\documentclass[10pt]{article}

\usepackage{amsfonts,amssymb,amsmath,amscd,latexsym}
\usepackage{doi}

\usepackage{filecontents}
\usepackage[numbers]{natbib}
\usepackage{amsthm}
\usepackage{mathtools}
\usepackage{tikz}
\usepackage{amsthm}

\usepackage{hyperref}
\hypersetup{
   colorlinks   =  true,
    citecolor    = red,
     urlcolor	=magenta,     
}

\newcommand{\sect}[1]{\setcounter{equation}{0}\section{#1}}
\newcommand{\subsect}[1]{\subsection{#1}}

\renewcommand{\theequation}{\arabic{section}.\arabic{equation}}

\def\be{\begin{equation}}
\def\ee{\end{equation}}
\def\bea{\begin{eqnarray}}
\def\eea{\end{eqnarray}}

\def\1{\'{\i}}                           
 
\def\back{\!\!\!\!\!\!}

\def\pa{\alpha}
\def\pb{\beta}

\def\c{\Lambda}%cosmological constant%

\DeclareMathOperator{\spn}{span}

\begin{document}
 
\begin{center}
\baselineskip 24 pt {\LARGE \bf  
The Poincar\'e group as a Drinfel'd double} 
\end{center}

\bigskip 
\medskip

\begin{center}

{\sc Angel Ballesteros$^1$, Ivan Gutierrez-Sagredo$^1$, Francisco J. Herranz$^1$}

{$^1$ Departamento de F\1sica, Universidad de Burgos, 
09001 Burgos, Spain}

e-mail: {\href{mailto:angelb@ubu.es}{angelb@ubu.es}, \href{mailto:igsagredo@ubu.es}{igsagredo@ubu.es}, \href{mailto:fjherranz@ubu.es}{fjherranz@ubu.es}}

\end{center}

\begin{abstract}
The eight nonisomorphic Drinfel'd double (DD) structures for the Poincar\'e Lie group in (2+1) dimensions are explicitly constructed in the kinematical basis. Also, the two existing DD structures for a non-trivial central extension of the (1+1) Poincar\'e group are also identified and constructed, while in (3+1) dimensions no Poincar\'e  DD structure does exist. Each of the DD structures here presented has an associated canonical quasitriangular Poincar\'e $r$-matrix whose properties are analysed. Some of these $r$-matrices give rise to coisotropic Poisson homogeneous spaces with respect to the Lorentz subgroup, and their associated Poisson Minkowski spacetimes are constructed. Two of these (2+1) noncommutative DD Minkowski spacetimes turn out to be quotients by a Lorentz Poisson subgroup: the first one corresponds to the double of $\mathfrak{sl}(2)$ with trivial Lie bialgebra structure, and the second one gives rise to a quadratic noncommutative Poisson Minkowski spacetime. With these results, the explicit construction of DD structures for all Lorentzian kinematical groups in (1+1) and (2+1) dimensions is completed, and the connection between (anti-)de Sitter and Poincar\'e $r$-matrices through the vanishing cosmological constant limit is also analysed.
\end{abstract}

\medskip 
\medskip 


\medskip 


%%%%%%%%%%%%%%%%%%%%%%%%%%%%%%%%%%%%%%%%%%%%%%%

\sect{Introduction} 
\label{Introduction}

The relation between quantum groups and quantum gravity has been suggested from different viewpoints since the introduction of the former by Drinfel'd \cite{Drinfeld1987icm} more than three decades ago. In fact, quantum group symmetries have arised in different approaches to quantum gravity, such as the combinatorial quantization of gravity in (2+1) dimensions or path integral approaches like state sum models or spin foams (see, for instance, \cite{TV1992invariants,AGS1995,AGS1996,AS1996duke,BR1995,MS2003,NR2003cosmological,FM2012}, and references therein). From a different point of view, quantum groups have also been introduced in order to describe the phenomenology of spacetimes in which the relativistic kinematical symmetries are modified by a Planck scale deformation. These approaches include noncommutative spacetime models like the one arising from $\kappa$-deformations of the Poincar\'e algebra (see, for instance,~\cite{Majid1988,LRNT1991,LNR1992fieldtheory,BHOS1995nullplane,BRH2003minkowskian}), doubly special relativity theories \cite{Amelino-Camelia2010symmetry,LN2003versus,FKS2004gravity} and related phenomena like non-trivial momentum space \cite{Kowalski-Glikman2013living,GM2013relativekappa,BGGH2017curvedplb,BGGH31}.

While in any dimension quantum groups are natural candidates to describe the symmetries of a quantum theory of gravity, in the case of (2+1) dimensions their role is much better understood. As it is well-known, gravity in (2+1) dimensions is quite different from the full (3+1)-dimensional theory \cite{DJH1984,Carlip2003book}. The source of this difference can be traced back to the fact that in three dimensions the Ricci tensor completely determines the Riemann tensor. Therefore, every solution of the vacuum Einstein field equations is locally isometric to one of the three maximally symmetric spacetimes of constant curvature, (anti-)de Sitter or Poincar\'e, depending only on the value of the cosmological constant \cite{Carlip2003book}. As a consequence, gravity in (2+1) dimensions is a topological theory in which gravitational waves do not exist. In fact, (2+1)-dimensional gravity admits a description as a Chern-Simons theory with gauge group given by the group of isometries of the corresponding spacetime model~\cite{AT1986,Witten1988}. In this context the phase space structure of (2+1)-gravity is related with the moduli space of flat connections on a Riemann surface whose symmetries are given by certain Poisson-Lie (PL) groups \cite{AM1995moduli,FR1999moduli}, which are known to be the semiclassical counterpart of quantum groups. The relevant Poisson structure on this space admits a natural description in terms of coboundary Lie bialgebras associated with the gauge group. It is the presence of these PL groups playing the role of classical symmetries what makes clearer how quantum groups should enter in the game. 

Given the above considerations, while the generic role of PL and quantum groups in (2+1)-gravity is clear, the question of which quantum deformations are the relevant ones from the physical viewpoint is a matter of intense investigation~\cite{ASS2004,MS2009generalized,MS2009semidualization,BHM2014tallinn,BGHMN2017dice,MercatiSergola2018constraints}. In this context, both Lorentzian and Euclidean groups have been considered \cite{MS2003,MS2009generalized,MS2009semidualization,PW1990,BM1998topological,BNR2002,MN2010hilbert,MS2008quaternionic}. Moreover, there is evidence that relevant quantum deformations are the ones coming from a classical $r$-matrix arising from a Drinfel'd double (DD) structure, since this ensures that the Fock-Rosly condition is fulfilled, thus allowing a consistent definition of the Poisson structure on the moduli space of flat connections (see~\cite{BHM2013cqg} and references therein). These works made evident that a systematic study of all the possible DD structures for the isometry groups of  spacetime models for (2+1)-gravity was needed. However, while the DD structures for the (anti-)de Sitter groups have been fully described \cite{BHM2013cqg,BHM2010plb,BHM2014plb,BHMN2014sigma}, that is not the case for DD structures for the Poincar\'e group, thus preventing the complete understanding of Lorentzian DD structures and their relationships under Lie bialgebra contraction procedures~\cite{BGHOS1995quasiorthogonal}, for which the inclusion of the cosmological constant $\c$ as an explicit parameter has been proven to be very useful. 

In this context, another directly related problem is the construction of Poisson (resp.~quantum) Lorentzian spacetimes in such a way that their covariance properties can be appropriately generalized when the kinematical groups are promoted to PL (resp.~quantum) ones. This implies the characterization and explicit construction of Poisson Lorentzian spacetimes, and it turns out that among all posible PL structures on the kinematical group $G$, only some of them will allow the definition of the associated Poisson Lorentzian spacetime $M=G/H$ through the canonical projection $G\to M$. 
Within this framework, to impose the stabilizer $H$ of a point of the homogeneous space to be a Hopf subalgebra ({\em i.e.} a Poisson/quantum subgroup) turns out to be too restrictive \cite{Dijkhuizen1994} and the coisotropy condition for $H$  turns out to be the relevant one (see~\cite{BMN2017homogeneous} and references therein). Moreover, such coisotropy condition admits a precise characterization at the Lie bialgebra level, which is helpful in order to characterise physically relevant quantum deformations, as it has been recently shown in~\cite{BMN2017homogeneous,BM2018extended}.

With all the above in mind, the main objectives of this paper are three-fold:
\begin{itemize}
\item Firstly, to fill the gap concerning Lorentzian DD structures by  constructing explicitly the full set of DD structures for the (2+1)-dimensional Poincar\'e group. This will be based on the classifications given in~\cite{Gomez2000} and~\cite{SH2002}, and thus completing in this way the previous works \cite{BHM2013cqg,BHM2010plb,BHM2014plb,BHMN2014sigma} in which all the DD structures for the (2+1) (anti-)de Sitter Lie algebras have been presented. The connection between the latter results and the Poincar\'e DD structures here presented will  also be  analysed through Lie bialgebra contraction techniques.
\item To construct the Poisson Minkowski spacetimes corresponding to the five coisotropic Lie bialgebras that come from the Poincar\'e DD structures that we have previously obtained. Two of them will be of Poisson subgroup type, and the features of their associated noncommutative Minkowski spacetimes will be analysed.
\item Finally, to address the (1+1)-dimensional case by enlarging the (1+1) Poincar\'e algebra with a (non-trivial) central generator in order to have an extended even-dimensional Lie algebra. Surprisingly enough, this extended algebra can be endowed with two different DD structures, whose Poisson Minkowski spacetimes are also constructed. This completes the study of DDs for the Poincar\'e group, since it is well-known that in (3+1) and higher dimensions the Poincar\'e Lie algebra does not admit any DD structure due to the lack of a nondegenerate symmetric bilinear form, which is essential for the definition of an appropriate pairing. 
\end{itemize}

The structure of the paper is the following. In the next section we summarize all the basic background material needed for the paper: the notion of DD Lie algebra and its connection with Lie bialgebra structures, PL groups, Poisson homogeneous spaces, quantum groups and quantum spaces. Section 3 is devoted to present a first DD structure for the (2+1) Poincar\'e group which comes from the trivial Lie bialgebra structure of the $\mathfrak{sl}(2,\mathbb{R})$ algebra, which is the algebraic structure underlying the noncommutative spacetime arising in (2+1) Lorentzian quantum gravity with vanishing cosmological constant (see~\cite{BMS2002,MW1998}; in the Euclidean case, the corresponding DD  structure was studied in~\cite{MS2009semidualization, BM2003, Majid2005time, JMN2009}). This example is described in full detail, including the explicit construction of the PL structure for the Poincar\'e group which is associated with this DD. The corresponding Poisson homogeneous Minkowski spacetime is also presented, which turns out to be linear and  isomorphic to the $\mathfrak{so}(2,1)$ Lie algebra. In section 4 we recall the complete classification of DD structures for the (2+1) Poincar\'e algebra given in~\cite{Gomez2000,SH2002}, where  eight nonisomorphic cases are shown to exist. For each of them we provide in section 5 a change of basis to the kinematical generators of the (2+1) Poincar\'e algebra, and we relate our results with the complete classification of classical $r$-matrices for the (2+1) Poincar\'e algebra given in~\cite{Stachura1998}, which is summarized in the Appendix. Furthermore, we   present the kinematical version of the associated canonical Poincar\'e $r$-matrices, and the Poisson Minkowski spacetimes associated with the five $r$-matrices fulfilling the coisotropy condition are explicitly constructed. In section 6 we study the contraction scheme that provides those Poincar\'e DD $r$-matrices that can be obtained as the vanishing cosmological constant limit $\c \to 0$ of the (anti-)de Sitter DD $r$-matrices given in~\cite{BHM2013cqg}. By following the same pattern, section 7 presents the two possible DD structures for the one-dimensional central extension of the (1+1) Poincar\'e algebra, together with its associated $r$-matrices and extended Poisson homogeneous Minkowski spacetimes. Finally, a concluding section closes the paper, where we emphasize that the plurality of DD structures makes the (2+1) Poincar\'e algebra a quite distinguished one among all kinematical groups. The problem of finding DD structures for other kinematical algebras and other dimensions (including some non-existence results), as well as several open research lines are also commented.

%%%%%%%%%%%%%%%%%%%%%%%%%%%%%%%%%%%%%%%%%%%%%%%

\sect{From Drinfel'd doubles to Poisson homogeneous spaces} \label{DD_Lie_algebras}

%%%%%%%%%%%%%%%%%%%%%%%%%%%%%%%%%%%%%%%%%%%%%%%

\subsect{Drinfel'd double structures}

A $2d$-dimensional Lie algebra $\mathfrak{a}$ is said to be the Lie algebra of  a (classical) DD Lie group~\cite{Drinfeld1987icm}   if there exists a basis $\{Y_1,\dots,Y_d,y^1,\dots,y^d \}$ of $\mathfrak a$ in which the 
Lie bracket reads
\begin{align}
[Y_i,Y_j]= c^k_{ij}Y_k, \qquad  
[y^i,y^j]= f^{ij}_k y^k, \qquad
[y^i,Y_j]= c^i_{jk}y^k- f^{ik}_j Y_k  \, .\label{agd}
\end{align}
Therefore, $\mathfrak{g}=\mbox{span}\{Y_1,\dots,Y_d\}$ and $\mathfrak{g}^\ast=\mbox{span}\{y^1,\dots,y^d \}$ define two Lie subalgebras with structure constants  $c^k_{ij}$ and $f^{ij}_k$, respectively. The triple $(\mathfrak{g},\mathfrak{g}^\ast,\mathfrak{a})$ is called a Manin triple, and the (classical) DD is the unique connected and simply connected Lie group $G$ with Lie algebra given by $\mathfrak{a}$. The Lie algebra
$\mathfrak{a}$ is said to be the double Lie algebra of $\mathfrak{g}$ and of its dual Lie algebra $\mathfrak{g}^\ast$. This duality is defined with respect to the nondegenerate symmetric bilinear form $\langle \, , \, \rangle : \mathfrak{a} \times \mathfrak{a} \rightarrow \mathbb{R}$ given by
\begin{align}\label{ages}
 \langle Y_i,Y_j\rangle= 0,\qquad \langle y^i,y^j\rangle=0, \qquad
\langle y^i,Y_j\rangle= \delta^i_j,\qquad \forall i,j  ,
\end{align}
which is `associative' or invariant in the sense that
\begin{align}\label{associative}
\langle [X,Y],Z \rangle = \langle X,[Y,Z] \rangle,  \qquad \forall X,Y,Z \in \mathfrak{a}. 
\end{align}
For any pair of structure constant tensors  $c^k_{ij}$ and $f^{ij}_k$, the universal enveloping algebra $U(\mathfrak{a})$ of a double Lie algebra has the following quadratic Casimir element 
\begin{align}
C=\tfrac12\sum_{i}{(y^i\,Y_i+Y_i\,y^i)},
\label{cascas}
\end{align}
which is directly related to the bilinear form~\eqref{ages}. Note that $U(\mathfrak{a})$ may of course have other Casimir elements in addition to $C$.

A given even-dimensional Lie group $G$ with Lie algebra $\mathfrak{a}$ can have several DD structures or no DD structures at all, {\em i.e.} several splittings of $\mathfrak{a}$ of the form~\eqref{agd} that are nonisomorphic in the sense that they cannot be transformed among them by a change of basis that preserves the full DD structure, including the pairing~\eqref{ages} (see~\cite{SH2002} for the complete classification of real 6-dimensional DD structures). Note that the DD  of $\mathfrak{g}$ and $\mathfrak{g}^\ast$ is always  isomorphic to the DD of $\mathfrak{g}^\ast$ and $\mathfrak{g}$.

The existence of a nondegenerate symmetric bilinear form~\eqref{ages} for a given even-dimensional Lie algebra is thus a necessary condition for the existence of DD structures. As it is well-known, the (3+1) Poincar\'e algebra does not have such a nondegenerate bilinear form (see, for instance,~\cite{Figueroa-OFarrill2018}), a fact that precludes the existence of DD structures for this group. On the contrary, such a nondegenerate pairing does exist for both the (2+1) Poincar\'e algebra and for the centrally extended (1+1) Poincar\'e algebra, and their explicit construction and analysis will be the main objective of the paper.

%%%%%%%%%%%%%%%%%%%%%%%%%%%%%%%%%%%%%%%%%%%%%%%

\subsect{Lie bialgebras and Drinfel'd double structures}

A Lie bialgebra ($\mathfrak{g},\delta$) is a Lie algebra $\mathfrak{g}$ endowed with a skew-symmetric cocommutator map
$
\delta:\mathfrak{g}\to \mathfrak{g}\otimes \mathfrak{g},
$
fulfilling the two following conditions:
\begin{itemize}
\item i) $\delta$ is a 1-cocycle, {\em  i.e.},
\be
\delta([X,Y])=[\delta(X),\,  Y\otimes 1+ 1\otimes Y] + 
[ X\otimes 1+1\otimes X,\, \delta(Y)] ,\qquad \forall \,X,Y\in
\mathfrak{g}.
\label{1cocycle}
\ee
\item ii) The dual map $\delta^\ast:\mathfrak{g}^\ast\otimes \mathfrak{g}^\ast \to
\mathfrak{g}^\ast$ is a Lie bracket on $\mathfrak{g}^\ast$.
\end{itemize}
It turns out that DD Lie algebras are in one-to-one correspondence with Lie bialgebra structures ($\mathfrak{g},\delta$) on $\mathfrak{g}$, by endowing the first Lie subalgebra $\mathfrak{g}$ with a skew-symmetric cocommutator map defined by the structure constants for $\mathfrak{g}^\ast$, namely
\be
[Y_i,Y_j]= c^k_{ij}Y_k, \qquad  \delta(Y_n)=f_{n}^{lm} Y_l\otimes Y_m.
\label{qqa}
\ee
In fact, the Jacobi identity for the DD Lie algebra~\eqref{agd} is just the 1-cocycle condition~\eqref{1cocycle} for the cocommutator map $\delta$ with respect to the Lie algebra $\mathfrak{g}$, which provides the following compatibility equations among the structure constants $c^k_{ij}$ and $f^{lm}_n$:
\be
f^{lm}_k c^k_{ij} = f^{lk}_i c^m_{kj}+f^{km}_i c^l_{kj}
+f^{lk}_j c^m_{ik} +f^{km}_j c^l_{ik}  \, . \label{compat}
\ee

Moreover, each DD structure for a Lie algebra $\mathfrak{a}$ generates a solution $r$ of the classical Yang-Baxter equation (CYBE) on $\mathfrak{a}$, which is provided by the canonical  classical
$r$-matrix
\be
r=\sum_i{y^i\otimes Y_i} .
\label{rcanon}
\ee
Therefore, if $\mathfrak{a}$ is a finite-dimensional DD Lie algebra,
then it  can be endowed with the
(quasitriangular) coboundary Lie bialgebra structure $(\mathfrak{a},\delta_D)$ 
 given by
\be
\delta_D(X)=[ X \otimes 1+1\otimes X ,  r],
\quad
\forall X\in {\mathfrak{a}}.
\label{rcanon2}
\ee
Explicitly, 
\be
\delta_D(y^i)=c^i_{jk}\,y^j\otimes y^k ,\qquad
\delta_D(Y_i)= - f_i^{jk}\,Y_j\otimes
Y_k \, .
\label{codob}
\ee
It is worth remarking that  $(\mathfrak{g},\delta_D)$ and ($\mathfrak{g}^\ast
,\delta_D$) are sub-Lie bialgebras of $(\mathfrak{a}, \delta_D)$. It is also important to stress  that  the cocommutator $\delta_D$  
is skew-symmetric, and therefore it depends only on the  skew-symmetric component $r'$ of the $r$-matrix (\ref{rcanon}), given by
\be  r'=\tfrac12 \sum_i{y^i\wedge Y_i}  .
\label{rmat}
\ee
This is explained by the fact that the symmetric component $\Omega$ of the $r$-matrix is just the tensorized form of the quadratic Casimir element of  $\mathfrak a$
(\ref{cascas}),
namely\be\label{omega}
\Omega=r-r'=\tfrac12\sum_i{(y^i\otimes Y_i+Y_i\otimes y^i)},
\ee
and we have that
\be [ Y
\otimes 1+1\otimes Y ,\Omega]=0 ,\qquad\forall Y\in \mathfrak a.
\ee

Expression~\eqref{omega} is just the Fock-Rosly condition for the $r$-matrix defining a PL structure on the extended phase space of holonomies in order to be compatible with the requirements of a Chern-Simons (2+1) gravity theory with gauge group given by the Lie group associated with $\mathfrak a$ (see~\cite{FR1999moduli,MS2003poisson,OS2018} and references therein). Therefore, DD structures for the Lorentzian kinematical groups (Poincar\'e and (anti-)de Sitter) play a relevant role in (2+1) gravity coupled to point particles.

%%%%%%%%%%%%%%%%%%%%%%%%%%%%%%%%%%%%%%%%%%%%%%%

\subsect{Poisson-Lie groups and Poisson homogeneous spaces}

For a coboundary PL group $G$, its Poisson structure $\Pi$ on $G$ is given by the so-called Sklyanin bracket \cite{ChariPressley}
\be
\{f,g\}= r^{ij} \left( \nabla^L_i f \nabla^L_j g - \nabla^R_i f \nabla^R_j g \right), \qquad f,g \in \mathcal C ^\infty(G),
\label{sklyanin}
\ee
which is written in terms of the components $r^{ij}$ of the classical $r$-matrix and the left- and right-invariant vector fields $\nabla^L_i,\nabla^R_i$ on the Lie group $G$. A well-known result by Drinfel'd \cite{Drinfeld1983hamiltonian} ensures that PL structures on a simply connected Lie group $G$ are in one-to-one correspondence with Lie bialgebra structures $\left(\mathfrak{g},\delta\right)$ on $\mathfrak{g}=\mbox{Lie}\left(G\right)$, and in the above coboundary case $\delta( {X} ) = \left(ad_{X} \otimes 1 + 1 \otimes ad_{X} \right) \left(r\right)$, where $r\in \mathfrak{g}\wedge \mathfrak{g}$ is a solution of the modified CYBE.

A Poisson homogeneous space (PHS) of a PL group $(G,\Pi)$ is a Poisson manifold $(M,\pi)$ endowed with a transitive group action $\rhd: G\times M\to M$ that is a Poisson map with respect to the Poisson structure on $M$ and the product $\Pi \times \pi$ of the Poisson structures on $G$ and $M$. In particular, let us consider a given spacetime $M$ as a  homogeneous space $G/{H}$ generated by a kinematical group $G$ and the corresponding isotropy subgroup $H$, whose Lie algebra will be denoted by $\mathfrak{h}$. A Poisson homogeneous spacetime $(M,\pi)$ is constructed by endowing the kinematical group $G$ with a PL structure $\Pi$~\eqref{sklyanin}, and the spacetime $M$ 
with a Poisson bracket $\pi$ 
that has to be compatible with the group action $\rhd: G\times M\to M$ in the abovementioned sense.

We stress that, given a spacetime $M=G/H$, a plurality of PHSs can be defined, since $G$ admits several PL structures $\Pi$ (which is tantamount to say that $\mathfrak g$ admits several Lie bialgebra structures $(\mathfrak g,\delta)$). Their characterization was given by  Drinfel'd, who proved that each PHS is in one-to-one correspondence with a Lagrangian subalgebra of the double Lie algebra $D(\mathfrak g)$ associated with the cocommutator $\delta$ \cite{Drinfeld1993}. A distinguished subclass of PHSs is the one in which the bracket $\pi$ can be just obtained as the canonical projection onto $M$ of the PL bracket $\Pi$. At the Lie bialgebra level, it can be shown (see~\cite{BMN2017homogeneous} for details) that this statement is equivalent to imposing the so-called coisotropy condition for the cocommutator $\delta$ with respect to the isotropy subalgebra $\mathfrak h$, namely
\be
\delta(\mathfrak h) \subset \mathfrak h \wedge \mathfrak g.
\label{coisotropy}
\ee
Furthermore, imposing the subgroup $H$ to have a sub-Lie bialgebra structure satisfying $\delta\left(\mathfrak{h}\right) \subset \mathfrak{h} \wedge \mathfrak{h}$ leads to a more restrictive constrain than the `true' coisotropic case~\eqref{coisotropy}, and implies that the PHS is constructed through an isotropy subgroup which is a Poisson subgroup with respect to $\Pi$.

%%%%%%%%%%%%%%%%%%%%%%%%%%%%%%%%%%%%%%%%%%%%%%%

\subsect{Quantum groups and quantum spaces}

Finally, we sketch the way in which DD structures and quantum deformations are linked through Lie bialgebras, since a given quantum universal enveloping algebra  $(U_z(\mathfrak g),\Delta_z)$ of a Lie algebra  $\mathfrak g$  is a Hopf algebra deformation of the universal enveloping algebra $(U(\mathfrak g),\Delta_0)$ `in the direction' of a certain Lie bialgebra 
$(\mathfrak g,\delta)$~\cite{ChariPressley}. Explicitly, the cocommutator map $\delta$ that specifies the deformation is given by the skew-symmetric part of the first-order of the coproduct  $\Delta_z$ in terms of the deformation parameter $z$, namely:
\be
\delta (X)=\frac 1 2\bigl(\Delta_z (X)-\sigma\circ \Delta_z (X)\bigr) +\mathcal O(z^2),\qquad
\forall X\in \mathfrak g ,
\ee
where $ \sigma$ is the flip operator  $ \sigma (X\otimes
Y)=Y\otimes X$. Moreover, if $(\mathfrak g,\delta)$ is a coboundary Lie bialgebra and $(U_z(\mathfrak g),\Delta_z)$ is a quasitriangular Hopf algebra, then the universal quantum $R$-matrix of the latter will be of the form $R=\mathbb I + z\, r + \mathcal O(z^2)$, where $r\in \mathfrak g\otimes \mathfrak g$ is a solution of the CYBE. Also, the dual Hopf algebra to $(U_z(\mathfrak g),\Delta_z)$ will be the corresponding quantum group $G_q$ (where $q=e^z$), which is just the quantization of the PL group $(G,\Pi)$ whose Poisson structure $\Pi$ is given by $r$ through the Sklyanin bracket~\eqref{sklyanin}. The quantization of a coisotropic Poisson homogeneous space $(M,\pi)$ will give rise to the quantum homogeneous space $M_q$, onto which the quantum group $G_q$ co-acts covariantly \cite{Dijkhuizen1994}.

All the previous ones are definitions for a generic Lie algebra $\mathfrak g$. When appropriate, and in order to emphasize the DD structure of the Lie algebra $\mathfrak g$, we will  denote $\mathfrak g\equiv \mathfrak a$, and the Lie bialgebra will be the canonical one $\delta_D$ coming from the DD structure. To summarise, if a Lie algebra $\mathfrak{a}$ has a DD structure (\ref{agd}),  then $(\mathfrak{a},\delta_D)$ is automatically a quasitriangular Lie bialgebra with canonical  $r$-matrix given by (\ref{rcanon}) and associated PL group given by $(G,\Pi_D)$. Therefore, the quantum algebra $(U_z(\mathfrak{a}),\Delta_z)$ has $\delta_D$ as its first-order deformation of the coproduct $\Delta_z$, and the PHS $(M,\pi_D)$ will be the semiclassical counterpart of the quantum homogeneous space induced by this quantum deformation. 

The aim of this paper is to obtain all the DD structures for the Poincar\'e Lie algebra by making use of the complete classifications of four- and six-dimensional real DDs given in~\cite{Gomez2000,SH2002}, and to construct explicitly their associated Poisson Minkowski spacetimes, which contain the essential information concerning the type of noncommutativity that can be expected for the corresponding quantum Minkowski spacetimes.

%%%%%%%%%%%%%%%%%%%%%%%%%%%%%%%%%%%%%%%%%%%%%%%

\sect{The (2+1) Poincar\'e algebra as $D(\mathfrak{sl}(2,\mathbb{R}))$} \label{sec:Dsl2}

To the best of our knowledge, the only DD structure for the (2+1) Poincar\'e algebra that has been studied so far in the literature is the `trivial' one, {\em i.e.} the one that comes from the trivial ($\delta=0$) Lie bialgebra structure of the three-dimensional $\mathfrak{sl}(2,\mathbb{R})$ algebra~\cite{BHM2014plb,MW1998} (for its Euclidean counterpart coming from $\mathfrak{su}(2)$ see~\cite{MS2009semidualization,BM2003, Majid2005time, JMN2009}). In this section we recall the construction in detail of this DD structure, which is usually called $D(\mathfrak{sl}(2,\mathbb{R}))$, and we will also provide all the technical aspects of the (2+1) Poincar\'e group (local coordinates and invariant vector fields) that will be needed in the rest of the paper.

%%%%%%%%%%%%%%%%%%%%%%%%%%%%%%%%%%%%%%%%%%%%%%%

\subsect{Kinematical basis and $r$-matrices}

It is well-known that all the Lie bialgebra structures for the (2+1) Poincar\'e algebra $\mathfrak{iso}(2,1)\equiv \mathfrak{p}(2+1)$ are coboundary ones~\cite{Zakrzewski1997}. The complete classification of nonisomorphic classes of $r$-matrices for  $\mathfrak{p}(2+1)$, which will be reviewed in the Appendix, is due to Stachura~\cite{Stachura1998} (recall that the (3+1) classification was done by Zakrzewski in~\cite{Zakrzewski1997}). In the kinematical basis we are going to work with, the commutation rules for the (2+1) Poincar\'e algebra are
\begin{align} 
\begin{array}{lll}
 [J,K_1]=K_2, & \quad [J,K_2]=-K_1,  & \quad  [K_1, K_2]=-J,  \\[2pt]
 [J,P_0]=0 ,& \quad [J,P_1]=P_2 ,& \quad [J, P_2]=-P_1 ,\\[2pt]
[K_1,P_0]=P_1 ,& \quad [K_1,P_1]=P_0 ,& \quad [K_1, P_2]=0,\\[2pt]
[K_2,P_0]=P_2 ,& \quad[K_2,P_1]= 0 ,& \quad[K_2, P_2]=P_0,\\[2pt]
[P_0,P_1]=0, & \quad[P_0, P_2]=0 ,& \quad  [P_1,P_2]=0 ,
 \end{array}
\label{aa}
\end{align}
where $\{J, K_1, K_2, P_0, P_1, P_2\}$ are, respectively, the generators of rotations, boosts, time translation and space translations. The two quadratic Casimir elements for this algebra are
\be
C_1=P_0^2 - P_1^2 - P_2^2,\qquad
C_2=\tfrac 12 \bigl( J\,P_0+P_0\,J +K_2\,P_1+P_1\,K_2 - ( K_1\,P_2 + P_2\,K_1) \bigr) .
\label{pcasimirs}
\ee

Minkowski spacetime in (2+1) dimensions $M^{2+1}$ thus arises as the homogeneous space of the Poincar\'e isometry group $ISO (2,1 )\equiv P(2+1)$ having the Lorentz subgroup $H=SO (2,1 )$ as the isotropy subgroup of the origin, namely $M^{2+1}\equiv  ISO (2,1 )/SO (2,1 )$. So we shall denote 
$$
\mathfrak{a}=\mathfrak{iso} (2,1 )=\mathfrak{p}(2+1)=\spn\{ J,K_1,K_2,P_0,P_1,P_2 \},\qquad \mathfrak{h}={\rm Lie}(H)=\mathfrak{so} (2,1 )=\spn\{ J,K_1,K_2 \}.
$$

A generic skew-symmetric element $r$  in $\mathfrak{p}(2+1)\wedge \mathfrak{p}(2+1)$ would be given by
\begin{eqnarray}
&& r = a_1 J \wedge P_1 + a_2 J\wedge K_1 + a_3 P_0\wedge P_1 + a_4 P_0 \wedge K_1 +a_5 P_1\wedge K_1 +a_6 P_1\wedge K_2 \cr 
 &&\qquad  + b_1 J\wedge P_2 + b_2 J\wedge K_2 + b_3 P_0\wedge P_2 + b_4 P_0\wedge K_2 +b_5 P_2\wedge K_2 + b_6 P_2\wedge K_1 \label{genericr}\\ 
 &&\qquad + c_1 J\wedge P_0 + c_2 K_1\wedge K_2 + c_3 P_1\wedge P_2 ,
\nonumber
\end{eqnarray}
where we have initially 15 possible `deformation parameters', which are constrained by the set of quadratic equations arising from the modified CYBE:
\be
[X\otimes 1\otimes 1 + 1\otimes X\otimes 1 +
1\otimes 1\otimes X,[[r,r]]\, ]=0, \qquad \forall X\in \mathfrak{p}(2+1),
\label{mCYBE}
\ee
where the Schouten bracket $[[r,r]]$ is defined as
\be
[[r,r]]:=[r_{12},r_{13}]+ [r_{12},r_{23}]+ [r_{13},r_{23}] ,
\label{Schouten}
\ee
in which $r_{12}=r^{ij}\,X_i \otimes X_j\otimes 1$, $r_{13}=r^{ij}\,X_i \otimes 1\otimes X_j$, $r_{23}=r^{ij}\,1 \otimes X_i\otimes X_j$. Recall that $[[r,r]]=0$ is just the  CYBE. A straightforward computation shows that the Schouten bracket for~(\ref{genericr})  reads 
\begin{eqnarray}
&& \back \back  [[r,r]]=(a_1 a_3 +a_4 b_3 + b_1 b_3 -a_3 b_4 - a_5 c_3 - b_5 c_3)   (P_1 \wedge K_1 \wedge P_0+P_1 \wedge P_2 \wedge J+ P_0\wedge P_2\wedge K_2)\cr
&&    \qquad+(a_4^2 +\! a_2 b_3 \!-\!  a_1 b_4 \!+a_6 c_1 \!+\! b_6 c_1 -\!a_5^2 \! -\! a_6 b_6 \!- \!c_2 c_3)\, (P_1 \! \wedge\! P_2\!\wedge\! P_0).
\label{schoutengen}
\end{eqnarray}
We recall that if $[[r,r]]=0$ we have triangular Lie bialgebras. Since all the $r$-matrices that we will obtain from DD structures are quasitriangular ones, their Schouten bracket will be non-vanishing.

%%%%%%%%%%%%%%%%%%%%%%%%%%%%%%%%%%%%%%%%%%%%%%%

\subsect{The `trivial' Drinfel'd double structure}

Let us consider the trivial $\delta=0$ Lie bialgebra structure for the $\mathfrak{g}=\mathfrak{sl}(2,\mathbb{R})$ algebra, which means that $\mathfrak{g}^\ast$ is the three-dimensional Abelian algebra. If we take the following basis for $ \mathfrak{sl}(2,\mathbb{R})$
\be
 [Y_0,Y_1]= 2 Y_1 ,
\qquad 
  [Y_0,Y_2]=  -2 Y_2 ,
\qquad 
  [Y_1,Y_2]= Y_0 ,
\ee  
together with a vanishing  cocommutator map  $\delta(Y_i)=0$, then the DD relations~\eqref{agd} lead to the 6-dimensional Lie algebra $\mathfrak a$ with brackets
\begin{align} 
\begin{array}{lll}
  [Y_0,Y_1]= 2 Y_1 ,&
\qquad 
  [Y_0,Y_2]=  -2 Y_2 ,&
\qquad 
  [Y_1,Y_2]= Y_0,
\\[2pt]
 [y^0,y^1]= 0 ,&
\qquad 
[y^0,y^2]=0, &
\qquad
 [y^1,y^2]=0 ,
 \\[2pt]
[y^0,Y_0]=0, &
\qquad 
 [y^0,Y_1]=y^2 ,&
\qquad 
[y^0,Y_2]=-y^1,
\\[2pt]
[y^1,Y_0]=2 y^1,&
\qquad 
 [y^1,Y_1]=-2 y^0 ,&
\qquad  
[y^1,Y_2]=0,
\\[2pt]
[y^2,Y_0]=- 2 y^2 ,&
\qquad 
[y^2,Y_1]= 0, &
\qquad 
[y^2,Y_2]=2 y^0.
 \end{array}
\label{pdoub}
\end{align}
The change of basis
\begin{align} 
\begin{array}{lll}
J=-\frac12 (Y_1 -Y_2) ,& \qquad
 K_1=\frac12 (Y_1 +Y_2), & \qquad
K_2=-\frac12  Y_0   , \\[2pt]
P_0=y^1 - y^2 , &\qquad
 P_1=2 y^0 ,&  \qquad
 P_2=y^1 + y^2,
  \end{array}
\label{csbasisp}
\end{align}
shows that this algebra is isomorphic to the (2+1) Poincar\'e algebra~\eqref{aa}, and the canonical pairing~\eqref{ages} for the kinematical generators is given by
\be
  \langle J,P_0\rangle=-1, \qquad \langle K_1,P_2\rangle=1, \qquad \langle K_2,P_1\rangle=-1 ,
  \label{pairing}
\ee
with all other entries equal to zero. Notice that the pairing is directly related  with the Casimir $C_2$ (\ref{pcasimirs}).

Therefore,  $\mathfrak{p}(2+1)$ can be thought of as a DD Lie algebra $\mathfrak a$, 
which under the inverse change of basis
\begin{align} 
\begin{array}{lll}
  Y_0= - 2 K_2, & \qquad
 Y_1=-J+K_1 ,&  \qquad
 Y_2=J+K_1   , \\[2pt]
y^0=\frac12 P_1, & \qquad
 y^1=\frac12 (P_0 +P_2), &  \qquad
 y^2=\frac12 (-P_0 +P_2) ,
  \end{array}
\label{csbasisinv3}
\end{align}
provides  the canonical classical $r$-matrix (\ref{rcanon}):
\be
r=\sum_{i=0}^2{y^i \otimes Y_i}=-P_0\otimes J - P_1\otimes K_2 + P_2\otimes K_1 .
\label{rcero}
\ee
By adding the tensorised Casimir $C_2$ (\ref{pcasimirs}),   the $r$-matrix can be skew-symmetrized and yields
\be
r'= \tfrac{1}{2} (- P_0 \wedge  J - P_1   \wedge K_2 + P_2  \wedge K_1 ),
\label{twists}
\ee
which is just    Class  (IV) in the Stachura classification~\cite{Stachura1998} of (2+1) Poincar\'e $r$-matrices (see   (\ref{classIV}) in  the Appendix where the translation of this classification in terms of the kinematical basis we are using throughout the paper is presented).

The $r$-matrix~\eqref{twists} is composed by three non-commuting twists,  and the DD structure above described induces a quantum Poincar\'e algebra whose cocommutator map induced by $r'$ is given by (\ref{rcanon2}) and reads
\bea
&&\delta_D(J)=\delta_D(K_1)=\delta_D(K_2)=0 ,\label{lorentz}\\[2pt]
 &&\delta_D(P_0)=  P_1\wedge P_2 ,\qquad \delta_D(P_1)= \, P_0\wedge P_2 ,\qquad \delta_D(P_2)= P_1\wedge P_0 .
\label{momenta}
\eea
Consequently, the corresponding quantum Poincar\'e algebra for which $\delta_D(J)$ gives the first-order deformation will have a non-deformed coproduct for the Lorentz sector, and the quantum deformation will be concentrated in the addition law for the translations sector.

It is also immediate to check that this Lie bialgebra is trivially coisotropic with respect to the Lorentz subalgebra $\mathfrak h=\mbox{span}\{ J,K_1,K_2 \}$, since $\delta_D(\mathfrak h)=0$ and the Lorentz subgroup is a (trivial) Poisson subgroup. Therefore, the canonical projection of the PL structure on $P(2+1)$ generated by $r'$~\eqref{twists} onto the Minkowski spacetime $M^{2+1}$  will give rise to a Poisson homogeneous Minkowski spacetime of Poisson subgroup type, whose quantisation will provide the noncommutative spacetime associated with this DD structure.

%%%%%%%%%%%%%%%%%%%%%%%%%%%%%%%%%%%%%%%%%%%%%%%

\subsection{An $\mathfrak{so}(2,1)$ noncommutative Minkowski spacetime}

In order to construct the Poisson bracket that defines the unique PL structure on the Poincar\'e group  $P(2+1)$ induced by the classical $r$-matrix $r'$ (\ref{twists}), we shall make use of the following  faithful matrix representation of the Poincar\'e Lie algebra, $\rho:{\mathfrak{p}(2+1)} \rightarrow \text{End} (\mathbb{R}^4)$:
\bea
&&\rho(J)= \begin{pmatrix}
0 & 0 & 0 & 0 \\ 
0 & 0 & 0 & 0 \\
0 & 0 & 0 & -1 \\
0 & 0 & 1 & 0 \\
\end{pmatrix} , \qquad
\rho(K_1) = \begin{pmatrix}
0 & 0 & 0 & 0 \\ 
0 & 0 & 1 & 0 \\
0 & 1 & 0 & 0 \\
0 & 0 & 0 & 0 \\
\end{pmatrix}  ,\qquad 
\rho(K_2) = \begin{pmatrix}
0 & 0 & 0 & 0 \\ 
0 & 0 & 0 & 1 \\
0 & 0 & 0 & 0 \\
0 & 1 & 0 & 0 \\
\end{pmatrix} ,\nonumber \\[4pt]
&& \rho(P_0) = \begin{pmatrix}
0 & 0 & 0 & 0 \\ 
1 & 0 & 0 & 0 \\
0 & 0 & 0 & 0 \\
0 & 0 & 0 & 0 \\
\end{pmatrix} , \qquad \ 
\rho(P_1) = \begin{pmatrix}
0 & 0 & 0 & 0 \\ 
0 & 0 & 0 & 0 \\
1 & 0 & 0 & 0 \\
0 & 0 & 0 & 0 \\
\end{pmatrix} , \qquad \ 
\rho(P_2) = \begin{pmatrix}
0 & 0 & 0 & 0 \\ 
0 & 0 & 0 & 0 \\
0 & 0 & 0 & 0 \\
1 & 0 & 0 & 0 \\
\end{pmatrix} .
\eea
By taking $\{\theta, \xi^1,\xi^2, x^0, x^1, x^2\}$ as the local coordinates on the group associated, in this order,  with the Lie generators $\{J, K_1, K_2, P_0, P_1, P_2\}$,   the Poincar\'e group element is given by
\be
G=\exp{(x^0 \,\rho(P_0))} \exp{(x^1 \, \rho(P_1))} \exp{(x^2 \, \rho(P_2))} \exp{(\xi^1 \, \rho(K_1))} \exp{(\xi^2 \, \rho(K_2))} \exp{(\theta \, \rho(J))},
\label{groupelement}
\ee
which can be straightforwardly computed and reads
\be
G=
\begin{pmatrix}
1 & 0 & 0 & 0 \\
x^0 & \cosh \xi^1 \cosh \xi^2 & \cos \theta \sinh \xi^1 + \sin \theta \cosh \xi^1 \sinh \xi^2 & \cos \theta \cosh \xi^1 \sinh \xi^2 - \sin \theta \sinh \xi^1 \\
x^1 & \sinh \xi^1 \cosh \xi^2 & \cos \theta \cosh \xi^1 + \sin \theta \sinh \xi^1 \sinh \xi^2 & \cos \theta \sinh \xi^1 \sinh \xi^2 - \sin \theta \cosh \xi^1 \\
x^2 & \sinh \xi^2 & \sin \theta \cosh \xi^2 & \cos \theta \cosh \xi^2
\end{pmatrix} .
\label{groupe}
\ee
Note that this set of coordinates is defined in such a way that $\{x^0, x^1, x^2\}$ are just the local coordinates on the homogeneous Minkowski spacetime $M^{2+1}$.

Now, the PL structure $\Pi_D$ on $P(2+1)$ is given by the Sklyanin bracket (\ref{sklyanin}) where in this case the $r^{ij}$ are the components of the skew-symmetric $r$-matrix $r '$ (\ref{twists}), and  $\nabla_i^{L (R)}$ denote the left (right) invariant vector fields on the Poincar\'e group, which turn out to be
\begin{align}
\begin{split}
&\nabla_J^L= \partial_\theta  ,\\
&\nabla_{K_1}^L=  \frac{\cos \theta}{\cosh \xi^2}  \left( \partial_{\xi^1} + \sinh \xi^2 \, \partial_\theta  \right) + \sin \theta \, \partial_{\xi^2} , \\
&\nabla_{K_2}^L= -\frac{\sin \theta}{\cosh \xi^2}  \left( \partial_{\xi^1} + \sinh \xi^2 \, \partial_\theta  \right) + \cos \theta \, \partial_{\xi^2} , \\
&\nabla_{P_0}^L=\cosh \xi^2 \left( \cosh \xi^1 \partial_{x^0} + \sinh \xi^1 \partial_{x^1} \right) + \sinh \xi^2 \partial_{x^2} , \\
&\nabla_{P_1}^L= \cos \theta  \left( \sinh \xi^1 \partial_{x^0} + \cosh \xi^1 \partial_{x^1} \right) + \sin \theta \left( \sinh \xi^2 \left( \cosh \xi^1 \partial_{x^0} + \sinh \xi^1 \partial_{x^1} \right) + \cosh \xi^2 \partial_{x^2} \right) ,\\
&\nabla_{P_2}^R= -\sin \theta  \left( \sinh \xi^1 \partial_{x^0} + \cosh \xi^1 \partial_{x^1} \right) + \cos \theta \left( \sinh \xi^2 \left( \cosh \xi^1 \partial_{x^0} + \sinh \xi^1 \partial_{x^1} \right) + \cosh \xi^2 \partial_{x^2} \right) ,
\end{split}
\end{align}
\begin{align}
\begin{split}
&\nabla_J^R= -x^2 \partial_{x^1} + x^1 \partial_{x^2} + \frac{\cosh \xi^1}{\cosh \xi^2} \left( \partial_\theta -\sinh \xi^2 \partial_{\xi^1} \right) + \sinh \xi^1 \partial_{\xi^2} , \\
&\nabla_{K_1}^R= x^1 \, \partial_{x^0} + x^0 \, \partial_{x^1} + \partial_{\xi^1} , \\
&\nabla_{K_2}^R= x^2 \partial_{x^0} + x^0 \partial_{x^2} + \frac{\sinh \xi^1}{\cosh \xi^2} \left( -\sinh \xi^2 \partial_{\xi^1} + \partial_\theta \right) + \cosh \xi^1 \partial_{\xi^2} , \\
&\nabla_{P_0}^R= \partial_{x^0} , \qquad
\nabla_{P_1}^R= \partial_{x^1}  ,\qquad
\nabla_{P_2}^R= \partial_{x^2} .
\end{split}
\end{align}

Therefore, the Poisson structure $\Pi_D$ is  found to be
\begin{align}
 \label{adspois}
 \{ x^0,  x^1\}= - \, x^2 , \qquad \{ x^0, x^2\}=  \,  x^1,  \qquad \{x^1, x^2\}=  \, x^0,
\end{align}
while the remaining Poisson brackets vanish. Hence, the canonical projection of the $\Pi_D$ brackets to the spacetime coordinates $ \{ x^0,  x^1, x^2\}$  gives rise to the Poisson Minkowski spacetime $\pi_D$ associated with this DD structure.
Thus, the relations (\ref{adspois}) define the Poisson Minkowski spacetime $({M}^{2+1},\pi_D)$, which is a Lie-algebraic Poisson spacetime isomorphic to the $\mathfrak{so}(2,1)$ algebra. By construction, this spacetime is covariant under the co-action defined by the Poincar\'e group element $G$~\eqref{groupe} and can be straightforwarly quantized, since no ordering ambiguities appear either in the Poisson bracket~\eqref{adspois} or in the coproduct induced by the group multiplication of two $G$ matrix elements. Note also that the Poisson brackets for the Lorentz coordinates vanish, in accordance with the fact that we have a trivial Lorentz Poisson subgroup with cocommutator $\delta_D(\mathfrak h)=0$.

This DD spacetime, together with its Euclidean $E^3$ counterpart giving rise to an $\mathfrak{so}(3)$ algebra, have been previously studied in the literature (see~\cite{MS2009semidualization,MW1998, BM2003, Majid2005time, JMN2009}). Both of them are Lie-algebraic spacetimes, and the representation theory of the corresponding algebra ($\mathfrak{so}(2,1)$ in the ${M}^{2+1}$ case and $\mathfrak{so}(3)$ in the $E^3$ one) characterises their physical properties. On the other hand, we recall that the very same DD construction applied to the Drinfel'd-Jimbo Lie bialgebra structure for $\mathfrak{sl}(2,\mathbb{R})$ was shown in~\cite{BHM2014plb} to give rise to a DD which is isomorphic to the (2+1)-dimensional anti-de Sitter algebra in which the deformation parameter $\eta$ defining a non-trivial Lie bialgebra structure $\delta$ is related to the cosmological constant in the form $\Lambda=-\eta^2$.

%%%%%%%%%%%%%%%%%%%%%%%%%%%%%%%%%%%%%%%%%%%%%%%

\sect{Nonisomorphic Drinfel'd double structures for $\mathfrak{p}(2+1)$} \label{otherDD}

The DD structure studied in the previous section is by no means the unique one, and the  three following sections will be devoted to the other seven DD structures that can be found for the (2+1) Poincar\'e algebra, together with their associated Poisson Minkowski spacetimes.
The existence of all these DD structures can be traced back to the work~\cite{Gomez2000}, where all the inequivalent three-dimensional real Lie bialgebra structures are classified, and to~\cite{SH2002} where all six-dimensional  nonisomorphic real DD structures were also classified. 

In the notation from~\cite{Gomez2000} we will be interested in the Lie bialgebra structures for three-dimensional real Lie algebras whose double Lie algebra $\mathfrak{a}$  is isomorphic to $\mathfrak{so}(2,1)\ltimes_{ad^*}\mathbb{R}^3\simeq\,\mathfrak{p}(2+1)$. There the cocommutator $\delta$~\eqref{qqa} for each three-dimensional Lie bialgebra is given by a classical $r$-matrix (which is denoted by $\chi$) for a given three-dimensional real Lie algebra  $\mathfrak g$, plus a non-coboundary contribution to the cocommutator which is denoted by $\tilde{\delta}$, namely
\be
\delta(e_a)=[1\otimes e_a + e_a \otimes 1, \chi] + \tilde{\delta}(e_a),
\qquad a=0,1,2.
\ee
In this notation, the DD structure described in the previous section (which is not included in \cite{Gomez2000} since the cases with trivial three-dimensional cocommutator are not considered) would be of the form
$\left( \mathfrak{g},\mathfrak{g^*} \right) = \left( \mathfrak{sl}_2,\text{Abelian}_3 \right)$  with $\chi = 0$, $\tilde{\delta} = 0$. Also, this case would correspond in~\cite{SH2002} to the DD denoted as $(8|1)$, and in the rest of the paper we will call it `Case 0'.

A very careful inspection and comparison of both classifications leads to the following seven additional non-trivial three-dimensional Lie bialgebra structures having $\mathfrak{p}(2+1)$ as double Lie algebra:
\begin{itemize}

\item \hyperref[sec:case1]{Case 1}: $\left( \mathfrak{g},\mathfrak{g^*} \right) = \left( \mathfrak{r}_3 (1),\mathfrak{sl}_2 \right) $  with $ \chi = \alpha e_0 \wedge e_1$, $\tilde{\delta} (e_1) = e_0 \wedge e_2 $ (which corresponds to Nr.~(3) from \cite{Gomez2000} and $f(8|5.iii)$ from \cite{SH2002}, where the notation $f()$ denotes the dual DD structure).

\item \hyperref[sec:case2]{Case 2}: $\left( \mathfrak{g},\mathfrak{g^*} \right) = \left( \mathfrak{r}_3 (1),\mathfrak{n}_3 \right) $  with $ \chi = 0$, $\tilde{\delta} (e_1) = e_0 \wedge e_2 $ (Nr.~10  from \cite{Gomez2000} and $(5|2.ii)$ from \cite{SH2002}).

\item \hyperref[sec:case3]{Case 3}: $\left( \mathfrak{g},\mathfrak{g^*} \right) = \left( \mathfrak{r}_3' (1),\mathfrak{n}_3 \right) $  with $ \chi =  0$, $\tilde{\delta} (e_2) = \lambda e_0 \wedge e_1$ (Nr.~13 from \cite{Gomez2000} and $(4|2.iii|b)$ from \cite{SH2002}). Note that in \cite{Gomez2000} there is a misprint stating that $\tilde{\delta} (e_1) = \lambda e_0 \wedge e_2$.

\item \hyperref[sec:case4]{Case 4}: $\left( \mathfrak{g},\mathfrak{g^*} \right) = \left( \mathfrak{s}_3 (0),\mathfrak{r}_3' (1) \right) $  with $ \chi =  \alpha e_0 \wedge e_1$, $\tilde{\delta} (e_0) = \lambda e_1 \wedge e_2$ (Nr.~(14')  from \cite{Gomez2000} and $(7_0 |4|b)$ from \cite{SH2002}).

\item \hyperref[sec:case5]{Case 5}: $\left( \mathfrak{g},\mathfrak{g^*} \right) = \left( \mathfrak{r}_3' (1),\mathfrak{r}_3 (-1) \right) $  with $ \chi =  \omega e_1 \wedge e_2$, $\tilde{\delta} (e_2) = \lambda e_0 \wedge e_1$ $(\omega \lambda >0)$ (Nr.~14  from \cite{Gomez2000} and $f(6_0 |4.i|b)$ from \cite{SH2002}). Note that in \cite{Gomez2000} there is a misprint stating that $\tilde{\delta} (e_1) = \lambda e_0 \wedge e_2$.

\item \hyperref[sec:case6]{Case 6}: $\left( \mathfrak{g},\mathfrak{g^*} \right) = \left( \mathfrak{r}_3 (1),\mathfrak{r}_3 (-1) \right) $ with $\chi = \omega e_1 \wedge e_2$ $(\omega > 0)$, $\tilde{\delta} (e_1)=e_0 \wedge e_2$ (Nr.~11 from \cite{Gomez2000} and $f(6_0 |5.i)$ from \cite{SH2002}).

\item \hyperref[sec:case7]{Case 7}: $\left( \mathfrak{g},\mathfrak{g^*} \right) = \left( \mathfrak{s}_3 (0),\mathfrak{r}_3 (1) \right) $  with $ \chi =  e_0 \wedge e_1$, $\tilde{\delta} = 0$ (Nr.~(11')  from \cite{Gomez2000} and $(7_0 |5.i)$ from \cite{SH2002}).

\end{itemize}

Thus, we have in total eight different DD structures whose commutation rules are displayed in Table~\ref{table: 2+1DD}. Notice that the double constructed from $\left( \mathfrak{g},\mathfrak{g^*} \right)$ is always  isomorphic to the one arising from $\left( \mathfrak{g^*},\mathfrak{g} \right)$. These eight DD structures for the (2+1) Poincar\'e algebra are nonisomorphic in the sense that for any pair of them there does not exist an algebra isomorphism that leaves the pairing~\eqref{ages} (and, therefore, the canonical classical $r$-matrix~\eqref{rcanon}) invariant. In the following section we will write all these DD structures in the kinematical basis~\eqref{aa}, where the expression of each canonical $r$-matrix will be different, and will fall into a given class within the Stachura classification described the Appendix. This reflects the fact that the inequivalence of 6-dimensional DD structures is translated into the inequivalence of the associated three-dimensional Lie bialgebra structures.

%%%%%%%%%%%%%%%%%%%%%%%%%%%%%%%%%%%%%%%%%%%%%%%

\begin{table}[tp]{\footnotesize
\caption{\small The eigth non-equivalent DD Lie algebras which are isomorphic to the (2+1) Poincar\'e algebra. The parameter $\omega$ can be rescaled to any non-zero real number of the same sign, while $\lambda$ is an essential parameter {different from zero}. In Case 5 they must obey $ \omega\lambda >0$. In Case 6 we have $\omega >0$.} \label{table: 2+1DD}
\begin{center}
\begin{tabular}{lllllllll}
\hline 
 &  &   &   &   &   &   &   &  \\[-1.5ex]
 & Case 0 & Case 1 & Case 2 & Case 3 & Case 4 & Case 5 & Case 6 & Case 7 \\[3pt]
\hline 
 &  &   &   &   &   &   &   &  \\[-1.5ex]
$[Y_0,Y_1]$ & $2 Y_1$ & $Y_1$ & $Y_1$ & $Y_1$ & ${-Y_2}$ & $Y_1$ & $Y_1$ & $-Y_2$\\[3pt] 
$[Y_0,Y_2]$ & $-2 Y_2$ & $Y_2$ & $Y_2$ & $Y_1+Y_2$ & ${Y_1}$ & $Y_1 + Y_2$ & $Y_2$ & $Y_1$\\[3pt]   
$[Y_1,Y_2]$ & $Y_0$ & $0$ & $0$ & $0$ & ${0}$ & $0$ & $0$ & $0$\\[3pt]  
\hline 
 &  &   &   &   &   &   &   &  \\[-1.5ex]
$[y^0,y^1]$ & $0$ & $y^0$ & $0$ & $\lambda y^2$ & ${0}$ & $\lambda y^2$ & $0$ & $0$ \\[3pt]   
$[y^0,y^2]$ & $0$ & $y^1$ & $y^1$ & $0$ & ${-y^0}$ & $0$ & $y^1$ & $-y^0$\\[3pt]   
$[y^1,y^2]$ & $0$ & $y^2$ & $0$ & $0$ & ${\lambda y^0 - y^1}$ & $2 \omega y^0$ & $2 \omega y^0$ & $-y^1$\\[3pt]  
\hline 
 &  &   &   &   &   &   &   &  \\[-1.5ex]
$[y^0,Y_0]$ & $0$ & $-Y_1$ & $0$ & $0$ & ${-Y_2}$ & $0$ & $0$ & $Y_2$\\[3pt]   
$[y^0,Y_1]$ & $y^2$ & $-Y_2$ & $-Y_2$ & $0$ & ${ \lambda Y_2-y^2 }$ & $0$ & $-Y_2$ & $0$\\[3pt]   
$[y^0,Y_2]$ & $-y^1$ & $0$ & $0$ & $- \lambda y^1$ & ${Y_0 - \lambda Y_1+y^1 }$ & $- \lambda Y_1$ & $0$ & $0$\\[3pt]   
$[y^1,Y_0]$ & $2 y^1$ & $Y_0+y^1$ & $y^1$ & $y^1+y^2$ & ${0}$ & $-2 \omega Y_2+y^1+y^2$ & $-2 \omega Y_2 + y^1$ & $y^2$\\[3pt]   
$[y^1,Y_1]$ & $-2 y^0$ & $-y^0$ & $-y^0$ & $-y^0$ & ${-Y_2}$ & $-y^0$ & $-y^0$ & $Y_2$\\[3pt]   
$[y^1,Y_2]$ & $0$ & $-Y_2$ & $0$ & $\lambda Y_0-y^0$ & ${Y_1-y^0}$ & $\lambda Y_0-y^0$ & $0$ & $-y^0$\\[3pt]   
$[y^2,Y_0]$ & $- 2 y^2$ & $y^2$ & $y^2$ & $y^2$ & ${0}$ & $2 \omega Y_1+y^2$ & $2 \omega Y_1 + y^2$ & $-Y_0-y^1$\\[3pt]   
$[y^2,Y_1]$ & $0$ & $Y_0$ & $Y_0$ & $0$ & ${y^0}$ & $0$ & $Y_0$ & $-Y_1+y^0$\\[3pt]   
$[y^2,Y_2]$ & $2 y^0$ & $Y_1-y^0$ & $-y^0$ & $-y^0$ & ${0}$ & $-y^0$ & $-y^0$ & $0$\\[3pt]  
\hline 
\end{tabular} 
\end{center}}
\end{table}

%%%%%%%%%%%%%%%%%%%%%%%%%%%%%%%%%%%%%%%%%%%%%%%

\sect{Drinfel'd double Poincar\'e $r$-matrices and Poisson Minkowski spacetimes} \label{PMS}

In the sequel we present, for each DD structure of the (2+1) Poincar\'e group given in Table~\ref{table: 2+1DD}, an invertible transformation to the kinematical basis (\ref{aa}) in which the canonical quasitriangular $r$-matrix~\eqref{rcanon} is given. Also, the Poisson homogeneous Minkowski spacetime arising from the DD $r$-matrices is explicitly computed for the five cases in which the Lie bialgebra is coisotropic with respect to the Lorentz subalgebra.

%%%%%%%%%%%%%%%%%%%%%%%%%%%%%%%%%%%%%%%%%%%%%%%

\subsection{Case 1}\label{sec:case1}

An isomorphism between the DD Lie algebra and  $\mathfrak{p}(2+1)$ in terms of the kinematical basis~\eqref{aa} reads
\begin{align}
&J=y^0 +y^1+y^2, & 
&K_1=y^0+y^1,&  
&K_2= -y^1-y^2 , \label{case1}\\
&P_0= y^0+y^1+Y_0-Y_1+Y_2 ,& 
&P_1= y^0+y^1+Y_0-Y_1,&  
&P_2= y^0-Y_1+Y_2.\nonumber
\end{align}
From the canonical pairing \eqref{ages}  we get the same~\eqref{pairing} up to a global sign:
\be
  \langle J,P_0\rangle=1,\qquad \langle K_1,P_2\rangle=-1, \qquad \langle K_2,P_1\rangle= 1.
  \label{xc}
\ee
By inserting the inverse of the basis transformation~\eqref{case1}    into~\eqref{rcanon}, the following classical $r$-matrix is found
\be
r_{1} =  \sum_{i=0}^2 y^i\otimes Y_i=   
K_1 \wedge J+ K_1\wedge K_2+J \otimes P_0 +K_2 \otimes P_1 -K_1\otimes P_2.
\label{r1}
\ee
And by subtracting the tensorized Casimir $C_2$~\eqref{pcasimirs}   from (\ref{r1}), one obtains the skew-symmetric $r$-matrix \begin{align} 
r'_{1} 
=K_1 \wedge J +K_1 \wedge K_2 +\bigl( - P_0 \wedge J - P_1 \wedge K_2 + P_2 \wedge K_1\bigr),
\label{r1skew}
\end{align}
which is a solution of the modified CYBE, that belongs to Class (I) in~\cite{Stachura1998}. In particular, if we apply the automorphism given by
\be
J\to J,\qquad K_1\to K_1,\qquad K_2\to K_2,\qquad P_i\to \sqrt{2} P_i, \quad i=0,1,2,
\ee
 to $r'_{1} $, we recover  (\ref{classI}) with $\alpha=1$ (up to a global constant $\sqrt{2}$). The cocommutator derived from~\eqref{rcanon2} is
\begin{align}
\begin{split}
\label{eq:cocom_Case1}
& \delta_D (J)= K_2 \wedge J,\\
& \delta_D (K_1)= J \wedge K_1+K_2 \wedge K_1    ,\\
& \delta_D (K_2)= J \wedge K_2,\\
& \delta_D (P_0)=J \wedge P_1 + P_2 \wedge K_1 + K_2 \wedge P_1 +2 P_1 \wedge P_2, \\
& \delta_D (P_1)= J \wedge P_0 + K_2 \wedge P_0 + P_2 \wedge K_1 + 2P_0 \wedge P_2  ,\\
& \delta_D (P_2)= P_0 \wedge K_1 + K_1 \wedge P_1+2 P_1 \wedge P_0.
\end{split}
\end{align}
Therefore, from the viewpoint of the construction of Poisson homogeneous Minkowski spacetimes, the Poincar\'e deformation induced by $r'_1$ is of Poisson subgroup type, since $\delta_D(\mathfrak h)\subset \mathfrak h \wedge \mathfrak h$ and the Lorentz subalgebra closes a sub-Lie bialgebra structure.

By making use of the results given in~\cite{BHM2014tallinn}, the DD $r$-matrix~\eqref{r1skew} can be thought of as a particular case of a more general solution of the modified CYBE that contains two independent real parameters $\pa_1,\pb_1$, namely
\begin{align}
r_{\rm 1, (\pa_1,\pb_1)} ' = \pa_1 \left(J \wedge K_1 +K_2 \wedge K_1\right) +\pb_1 \left( P_0 \wedge J+P_1 \wedge K_2 + K_1 \wedge P_2 \right),
\label{r1skewparam}
\end{align}
although only for the case with $\pa_1=\pb_1$ the DD structure is recovered. This embedding allows for a more clear interpretation of the contributions coming from each term of the $r$-matrix. In particular, the associated Poisson Minkowski spacetime can be obtained by computing the Sklyanin bracket for~\eqref{r1skewparam} and afterwards by projecting to the Poisson subalgebra generated by the spacetime coordinates $\{x^0,x^1,x^2\}$. A straightforward computation shows that the final result is
\begin{align}
\begin{split}
\label{eq:ncPst_Case1}
&\{ x^0,x^1 \}=-\pa_1 x^2 (x^0 + x^1) + 2 \pb_1 x^2 , \\
&\{ x^0,x^2 \}= \pa_1 x^1 (x^0 + x^1) - 2 \pb_1 x^1  ,\\
&\{ x^1,x^2 \}=\pa_1 x^0 (x^0 + x^1) - 2 \pb_1 x^0 ,
\end{split}
\end{align} 
which is a noncommutative {\em quadratic} Poisson Minkowski spacetime, whose linear part is ruled by the parameter $\pb_1$ (and is just proportional to the one in Case 0 (\ref{twists})) and comes from the dual of the cocommutator~\eqref{eq:cocom_Case1}. The quadratic part of the bracket comes from the $\pa_1$-contribution to the $r$-matrix, which is a triangular solution of the CYBE (a twist) of the type considered in~\cite{LW2006} in   (3+1) dimensions. Moreover, this twist is the one that generates the non-zero cocommutator for the Lorentz sector, thus being responsible of the Poisson subgroup structure of the isotropy subgroup. Hence, the Poisson Minkowski spacetime (\ref{eq:ncPst_Case1}) can be regarded as a quadratic generalization of the Lie-algebraic one (\ref{adspois}). As it is detailed in the Appendix, note that this is the only $r$-matrix for $\mathfrak{p}(2+1)$ having a term living in $\mathfrak{h} \wedge \mathfrak{h}$. We also remark that the quantization of this Poisson structure is far from being trivial, due to the ordering problems arising from the quadratic terms.

%%%%%%%%%%%%%%%%%%%%%%%%%%%%%%%%%%%%%%%%%%%%%%%

\subsection{Case 2}\label{sec:case2}

Now the Lie algebra  isomorphism  is given by
\begin{align}
&J=y^2+Y_0+Y_1, & 
&K_1=-y^2-Y_0,&  
&K_2= Y_0+Y_1, \label{case2}\\
&P_0=y^0-y^1-Y_2 ,& 
&P_1=-y^0+Y_2,& 
&P_2=-y^0+y^1,\nonumber
\end{align}
and the canonical pairing   is again (\ref{pairing}).  
The inverse of the basis transformation into~\eqref{rcanon} leads to
\be
r_{2} = P_2 \wedge J+ K_2 \wedge P_0 + K_2 \wedge P_2+ P_2 \otimes K_1 - P_1 \otimes K_2 - J \otimes P_0.
\label{r2}
\ee
This $r$-matrix can be straightforwardly skew-symmetrized through the tensorized Casimir $C_2$~\eqref{pcasimirs} yielding 
\begin{align} 
r'_{2} = P_2 \wedge J - P_0 \wedge K_2 - P_2 \wedge K_2 + \tfrac12 (P_0 \wedge J - P_1 \wedge K_2 + P_2 \wedge K_1),
\label{r2skew}
\end{align}
and it can be shown to belong to Class (IIa) in~\cite{Stachura1998} by applying the following automorphism to $r'_{2}$
\begin{align}
&  J \rightarrow - 2J -K_1 +\sqrt{2} \, K_2, && P_0 \rightarrow -2\bigl(2 P_0+\sqrt{2}\, P_1 + P_2\bigr), \nonumber \\
&  K_1 \rightarrow \bigl(1+\tfrac{1}{\sqrt{2}}\bigr) J + K_1 - \bigl(1+\tfrac{1}{\sqrt{2}}\bigr) K_2 , && P_1 \rightarrow   \bigl( 2-\sqrt{2}\bigr) P_0-\bigl(2-\sqrt{2}\bigr) P_1 + 2P_2, \nonumber \\
&  K_2  \rightarrow -\bigl( 1-\tfrac{1}{\sqrt{2}}\bigr) J - K_1 - \bigl(1-\tfrac{1}{\sqrt{2}}\bigr) K_2,  && P_2 \rightarrow \bigl(2+\sqrt{2}\bigr) P_0+\bigl(2+\sqrt{2}\bigr) P_1 + 2P_2,
\label{auto2}
\end{align}
which leads to the $r$-matrix (\ref{classIIa})    with parameters   $\rho=\alpha=1$ and   term $a=0$, namely
\begin{align} 
r'_{2} =  K_2  \wedge P_0+ J  \wedge P_1     - K_1 \wedge P_2 .
\label{r2skew2}
\end{align}
In this form, Case 2 can be clearly interpreted as the superposition of the `space-like' $\kappa$-Poincar\'e deformation \cite{BGHOS1995quasiorthogonal,BHOS1994global},  coming from the $r$-matrix $ K_2  \wedge P_0+ J  \wedge P_1$, along with a twist $K_1 \wedge P_2 $.

Next, by computing $\delta_D$, it can be shown that this DD structure generates a Poisson Minkowski spacetime fulfilling the coisotropy condition  $\delta_D(\mathfrak h)\subset \mathfrak h \wedge \mathfrak g$. By taking into into account~\cite{BHM2014tallinn}, we find that the embedding of \eqref{r2skew} into a more general solution of the modified CYBE is given by
\begin{align}
r_{\rm 2, (\pa_2,\pb_2)} ' = \pa_2 \left( P_0\wedge K_2+P_2\wedge K_2 \right)+\pb_2 \left( J \wedge P_2 +\tfrac{1}{2} \left( J\wedge P_0 + P_1 \wedge K_2 +  K_1 \wedge P_2 \right) \right),
\label{r2skewparam}
\end{align}
with $\pa_2,\pb_2\in \mathbb R$ (the DD case corresponds to   $\pa_2=\pb_2$).
The Poisson Minkowski spacetime turns out to be
\begin{align}
\label{eq:ncPst_Case2}
\{ x^0,x^1 \}=0 ,\qquad
\{ x^0,x^2 \}=-\pa_2 \left( x^0 - x^2 \right) , \qquad
\{ x^1,x^2 \}=-\pb_2 \left( x^0 - x^2 \right),
\end{align} 
which is linear and thus can be straightforwardly quantized.

%%%%%%%%%%%%%%%%%%%%%%%%%%%%%%%%%%%%%%%%%%%%%%%

\subsection{Case 3}\label{sec:case3}

The Lie algebra isomorphism   reads  
\begin{align}
\label{case3}
&J=\frac{1}{\lambda}(y^0+y^1)-Y_0+Y_2, & 
&K_1=-\frac{1}{\lambda} y^1 +Y_0-Y_1,&  
&K_2=-\frac{1}{\lambda} y^2 +Y_0-Y_2 , \\
&P_0= y^0+y^2+\lambda Y_1 ,& 
&P_1= y^0+\lambda Y_1,&  
&P_2= -y^0-y^2,\nonumber
\end{align}
and leads again to (\ref{xc}). The DD $r$-matrix is found to be
\begin{align}
& r_{3} =  J \wedge P_2 + K_2 \wedge P_0 + K_2 \wedge P_2  - P_2 \otimes K_1 + P_1 \otimes K_2 + J \otimes P_0 \cr
&\qquad\qquad\qquad  + \frac{1}{\lambda}\bigl(P_0 \wedge P_1 + 2 (P_0 \wedge P_2 + P_2 \wedge P_1)  + P_0 \otimes P_0 - P_1 \otimes P_1 - P_2 \otimes P_2\bigr),
\label{r3}
\end{align}
which by making use of the tensorised version of both Casimirs $C_1$ and $C_2$~\eqref{pcasimirs} can be transformed into:
\begin{align} 
& r'_{\rm 3} =
- P_2 \wedge J- P_0 \wedge K_2 - P_2  \wedge K_2  + \frac{1}{2}(- P_0 \wedge J + P_1 \wedge K_2 - P_2 \wedge K_1) \nonumber\\
&\qquad\qquad + \frac{1}{\lambda}\bigl(P_0 \wedge P_1 + 2 (P_0 \wedge P_2 + P_2 \wedge P_1)\bigr).
\label{r3skew}
\end{align}
This $r$-matrix  is shown to belong to Class (IIa)  in~\cite{Stachura1998} by applying to $ r'_{\rm 3} $ the composition of the automorphism (\ref{auto2}) and
\be
J\to -J,\qquad K_1\to -K_1,\qquad K_2\to K_2,\qquad P_0\to P_0,\qquad P_1\to -P_1,\qquad P_2\to P_2.
\label{auto2b}
\ee
In this way we obtain (\ref{classIIa}) with $\rho=\alpha=1$ but now with a term proportional to $1/\lambda$.
Hence  the difference between Cases 2 and 3 relies on the $1/\lambda$ term, which precludes the coisotropy condition $\delta_D(\mathfrak h)\subset \mathfrak h \wedge \mathfrak g$ to hold since, for instance,
\be
\delta_D(J)= P_0 \wedge  K_1  +P_1  \wedge  J + P_1 \wedge K_2 + P_2\wedge  K_1   + \frac{1}{\lambda} (P_0 \wedge P_2 + 2 P_1 \wedge P_0).
\ee
Therefore, this condition  is only fulfilled in the limit $\lambda\to \infty$, which leads to the previous Case 2.  Therefore the Poisson Minkowski spacetime for this case will not be constructed.

%%%%%%%%%%%%%%%%%%%%%%%%%%%%%%%%%%%%%%%%%%%%%%%

\subsection{Case 4}\label{sec:case4}

The isomorphism
\begin{align}
\label{case4}
&J= \lambda y^0 - Y_0, & 
&K_1= \lambda y^0 + y^1 + Y_0,&  
&K_2= y^2 +\lambda Y_2, \\
&P_0= y^0 - Y_1,& 
&P_1= -Y_2,&  
&P_2= Y_1,\nonumber
\end{align}
leads again to the (2+1) Poincar\'e algebra with  pairing (\ref{pairing}). The classical $r$-matrix is now
\begin{align}
& r_{\rm 4} =  P_2 \wedge J - J \otimes P_0 - P_1 \otimes K_2 + P_2 \otimes K_1 + \lambda (P_0 \otimes P_0 - P_1 \otimes P_1 - P_2 \otimes P_2 + P_0 \wedge P_2),
\label{r4}
\end{align}
which can be fully skew-symmetrized by making use of both Casimirs $C_1$ and  $C_2$~\eqref{pcasimirs} yielding
\begin{align} 
& r'_{4} =P_2 \wedge J + \tfrac12 (P_0 \wedge J - P_1 \wedge K_2 + P_2 \wedge K_1) + \lambda P_0 \wedge P_2.
\label{r4skew}
\end{align}
This $r$-matrix belongs to Class (IIIb) in~\cite{Stachura1998}, since $r'_{4} $ turns out to be proportional  to  the $r$-matrix (\ref{classIIIb})  with $\rho=1$ and term $a\not =0$  under the automorphism 
\be
J \rightarrow i K_2, \qquad K_1 \rightarrow i J, \qquad K_2 \rightarrow - K_1, \qquad  P_0 \rightarrow - i P_1, \qquad P_1 \rightarrow P_2, \qquad P_2 \rightarrow i P_0.
\label{auto4}
\ee
Again, due to the presence of the non-zero essential parameter $\lambda$ (i.e., $a\not =0$),    this case does not fulfil the coisotropy condition (\ref{coisotropy}).

%%%%%%%%%%%%%%%%%%%%%%%%%%%%%%%%%%%%%%%%%%%%%%%

\subsection{Case 5}\label{sec:case5}

Here we have two different subcases due to the constraint $\omega \lambda  >0$ (see Table~\ref{table: 2+1DD}): either $\omega > 0$ and $  \lambda>0$  or $\omega < 0$ and $\lambda < 0$. Although the isomorphism is different for each subcase, we shall show that the resulting $r$-matrices  are the same in both of them.

As stated in \cite{Gomez2000}, $\omega$ can be
rescaled to any non-zero value of the same sign and $\lambda$ is an essential parameter.
Therefore,  if we set $\omega=1/2$ and hence $\lambda > 0$, the isomorphism is given by
\begin{align}
\label{case51}
&J= \frac{1}{\sqrt{\lambda}}\, (-y^1+Y_1), & 
&K_1=-\frac{1}{\lambda}\,y^0-Y_0,&  
&K_2= \frac{1}{\sqrt{\lambda}} \,(y^1+Y_1 -Y_2),\\
&P_0=-\sqrt{\lambda}\,(y^2+Y_1) ,& 
&P_1=-\sqrt{\lambda} \,y^2,&  
&P_2=y^0,\nonumber
\end{align}
and the pairing is   (\ref{xc}).
On the other hand, if $\omega=-1/2$ and $\lambda < 0$, then
\begin{align}
\label{case52}
&J= -\frac{1}{\sqrt{-\lambda}} \,(y^1+Y_1), & 
&K_1=-\frac{1}{\lambda}\,y^0-Y_0,&  
&K_2= \frac{1}{\sqrt{-\lambda}}\,(y^1- Y_1 +Y_2), \\
&P_0= \sqrt{-\lambda} \,(y^2-Y_1) ,& 
&P_1= \sqrt{-\lambda} \, y^2,&  
&P_2= y^0,\nonumber
\end{align}
together with the same pairing   (\ref{xc}).

Both subcases lead to the same classical $r$-matrix  
\begin{align} 
& r_{5} =  P_1 \wedge J -P_2 \otimes K_1 +P_1 \otimes K_2 + J \otimes P_0+ \frac{1}{\lambda } \left(  P_1 \wedge P_0 + P_0 \otimes P_0 - P_1 \otimes P_1 - P_2 \otimes P_2  \right),
\label{r5}
\end{align}
which,  by making use of both Casimirs, can be written in  the skew-symmetric form
\begin{align} 
& r'_{5} = P_1 \wedge J + \frac{1}{2} \left( - P_0 \wedge J + P_1 \wedge K_2 - P_2 \wedge K_1 \right) + \frac{1}{\lambda} P_1 \wedge P_0,
\label{r5skew}
\end{align}
which is proportional to the $r$-matrix (\ref{classIIIb}) with $\rho=-1$ and term $a\not =0$, so belonging  to Class (IIIb) in~\cite{Stachura1998}. As in Case 4,  the parameter  $\lambda$ precludes the    coisotropy condition (\ref{coisotropy}) to be satisfied.

%%%%%%%%%%%%%%%%%%%%%%%%%%%%%%%%%%%%%%%%%%%%%%%

\subsection{Case 6}\label{sec:case6}

As the parameter $\omega > 0$ can be rescaled to any positive real number, hereafter we take $\omega=1/2$. An isomorphism between the kinematical basis and the DD one is given by 
\begin{align}
J&=Y_1 + y^2, & 
K_1&= Y_0, & 
K_2&= y^2, \label{isomcase6} \nonumber\\
P_0&= - y^1, &
P_1&= -Y_2 + y^1, &
P_2&= y^0 ,
\end{align}
with   pairing (\ref{pairing}).
The corresponding inverse isomorphism gives rise to the classical $r$-matrix
\begin{align}
& r_{\rm 6}  = P_0 \wedge K_2 + P_2 \otimes K_1 - K_2 \otimes P_1 - P_0 \otimes J,
\label{r6}
\end{align}
and with the aid of $C_2$ we obtain  
\begin{align} 
& r'_{6} = P_0 \wedge K_2 + \tfrac12 (- P_0 \wedge J + P_1 \wedge K_2 + P_2 \wedge K_1).
\label{r6skew}
\end{align}
By making use of the automorphism
\be
J \rightarrow J,  \qquad K_1 \rightarrow -K_1,    \qquad   K_2 \rightarrow - K_2, \qquad P_0 \rightarrow P_0,  \qquad P_1 \rightarrow -P_1,  \qquad P_2 \rightarrow -P_2,
\label{auto6}
\ee
we find that  $r'_{6}$ coincides (up to a factor 1/2) with   the $r$-matrix (\ref{classIIIb})  of Class (IIIb) with $\rho=1$ but now  with the term $a=0$. This is a coisotropic Lie bialgebra, whose Poisson Minkowski spacetime reads
\begin{align}
\begin{split}
\label{eq:ncPst_Case6}
&\{ x^0,x^1 \}=0 ,\qquad
 \{ x^0,x^2 \}=-x^0 + x^1 , \qquad
 \{ x^1,x^2 \}=0,
\end{split}
\end{align} 
and its quantization is   straightforward. 

Notice that Cases 4 and 6 are,  obviously, related since they are within the same Class  (IIIb) with parameter $\rho=1$. In fact, if we write $r'_4$ (\ref{r4skew}) under the automorphism (\ref{auto4}), its limit $\lambda\to 0$ leads to $r'_{6}$ (\ref{r6skew}) expressed under the map (\ref{auto6}).

%%%%%%%%%%%%%%%%%%%%%%%%%%%%%%%%%%%%%%%%%%%%%%%

\subsection{Case 7}\label{sec:case7}

Finally,  the kinematical and DD basis are now related through
\begin{align}
J&=y^0, & 
K_1&=-Y_2, & 
K_2&= -Y_1-y^0, \label{isomcase7} \nonumber\\
P_0&= Y_0-y^1, &
P_1&= -y^1, &
P_2&= y^2 ,
\end{align}
with pairing (\ref{xc}). The inverse isomorphism provides the classical $r$-matrix
\begin{align}
& r_{\rm 7} =    P_2 \wedge J + K_1 \otimes P_2 - K_2 \otimes P_1 - P_0 \otimes J .
\label{r7}
\end{align}
By subtracting $C_2$ we find  
\begin{align} 
& r'_{7} =P_2 \wedge J + \tfrac12 (- P_0 \wedge J + P_1 \wedge K_2 - P_2 \wedge K_1),
\label{r7skew}
\end{align}
which, under the automorphism
\be
J \rightarrow J, \qquad K_1 \rightarrow -K_2, \qquad K_2 \rightarrow K_1, \qquad P_0 \rightarrow P_0, \qquad P_1 \rightarrow -P_2, \qquad P_2 \rightarrow P_1,
\label{auto7}
\ee
turns out to  correspond  again to Case (IIIb) with  $r$-matrix (\ref{classIIIb}) such that $\rho=-1$ and the $a$ term vanishes. Note that $ r'_{7} $ (\ref{r7skew}), written under the automorphism (\ref{auto7}), is recovered from  $ r'_{5} $ (\ref{r5skew}) by taking the limit $\lambda\to \infty$. The coisotropy condition is fulfilled and the 
  associated Poisson  Minkowski spacetime is given by
\begin{align}
\begin{split}
\label{eq:ncPst_Case7}
\{ x^0,x^1 \}= 0,\qquad
\{ x^0,x^2 \}= 0,\qquad
\{ x^1,x^2 \}= - (x^0 + x^2),
\end{split}
\end{align} 
whose quantization is also straightforward.

%%%%%%%%%%%%%%%%%%%%%%%%%%%%%%%%%%%%%%%%%%%%%%%

\begin{table}[tp]{\footnotesize
\caption{\small  The (2+1) Poincar\'e $r$-matrices and Poisson subgroup/coisotropy condition for each of the eight DD structures on $\mathfrak{p}(2+1)$ as well as the corresponding class in the Stachura classification.} \label{table:2+1_r-matrices}\begin{center}
\begin{tabular}{cllc}
\hline 
 &  &   &       \\[-1.5ex]
Case &  Classical $r$-matrix  $r'_i$& $\delta_D \left(\mathfrak{h}\right)$ &Class~\cite{Stachura1998} \\[3pt]
\hline 
 &  &   &       \\[-1.5ex]
0 & $ \frac{1}{2} ( - P_0\wedge J - P_1 \wedge K_2 + P_2\wedge K_1)$ & $=0$& (IV) \\[5pt] 
1 & $K_1 \wedge J +K_1 \wedge K_2 +( - P_0 \wedge J - P_1 \wedge K_2 + P_2 \wedge K_1)$ & $\subset \mathfrak{h}\wedge \mathfrak{h}$ &  (I) \\[4pt] 
2 & $P_2 \wedge J - P_0 \wedge K_2 - P_2 \wedge K_2 + \frac12 (P_0 \wedge J - P_1 \wedge K_2 + P_2 \wedge K_1)$ & $\subset \mathfrak{h}\wedge \mathfrak{g}$ &  (IIa) \\[5pt]  
3 & $ - P_2 \wedge J - P_0 \wedge K_2 - P_2  \wedge K_2+ \frac{1}{2}(- P_0 \wedge J + P_1 \wedge K_2 - P_2 \wedge K_1)  $ & $\not\subset \mathfrak{h}\wedge \mathfrak{g}$   &(IIa) \\[4pt] 
  & $\qquad  + \frac{1}{\lambda}\bigl(P_0 \wedge P_1 + 2 (P_0 \wedge P_2 + P_2 \wedge P_1) \bigr)$ &    & \\[5pt] 
4 & {$P_2 \wedge J + \frac12 (P_0 \wedge J - P_1 \wedge K_2 + P_2 \wedge K_1) + \lambda P_0 \wedge P_2$} & {$\not\subset \mathfrak{h}\wedge \mathfrak{g}$} &(IIIb) \\[5pt]  
5 & $P_1 \wedge J + \frac{1}{2} \left( - P_0 \wedge J + P_1 \wedge K_2 - P_2 \wedge K_1 \right) + \frac{1}{\lambda} P_1 \wedge P_0$ & $\not\subset \mathfrak{h}\wedge \mathfrak{g}$ & (IIIb)\\[5pt]  
6 & $P_0 \wedge K_2 + \frac12 (- P_0 \wedge J + P_1 \wedge K_2 + P_2 \wedge K_1)$ & $\subset \mathfrak{h}\wedge \mathfrak{g}$ & (IIIb)\\[5pt]   
7 & $P_2 \wedge J + \frac12 (- P_0 \wedge J + P_1 \wedge K_2 - P_2 \wedge K_1)$ &  $\subset \mathfrak{h}\wedge \mathfrak{g}$ & (IIIb) \\[5pt] 
\hline
\end{tabular} 
\end{center}}
\end{table}

%%%%%%%%%%%%%%%%%%%%%%%%%%%%%%%%%%%%%%%%%%%%%%%

Therefore, by starting from the classification of Lie bialgebras given in \cite{Gomez2000}, we have obtained eight DD for the (2+1) Poincar\'e group. Table~\ref{table:2+1_r-matrices} summarizes our results: five of the classical $r$-matrices give rise to coboundary Lie bialgebras compatible with our algebraic conditions for $\delta_D (\mathfrak{h} )$ (\ref{coisotropy}), which guarantee that the Poisson bracket between Minkowski coordinates close a Poisson subalgebra. Among them, only Case 0 (trivially) and Case 1 turn out to be of Poisson subgroup type.

%%%%%%%%%%%%%%%%%%%%%%%%%%%%%%%%%%%%%%%%%%%%%%%

\sect{Contraction from (anti-)de Sitter  $r$-matrices} \label{StachDD}

Since the complete study of DD structures for the (anti-)de Sitter  (hereafter (A)dS) Lie algebras in (2+1) dimensions was performed in~\cite{BHM2013cqg}, it is therefore natural to study the behavior of the associated DD $r$-matrices  under the Lie algebra contraction to the Poincar\'e Lie algebra, that corresponds in kinematical terms to the $\c\to 0$ limit. Note that the classification of classical $r$-matrices for all real forms of $\mathfrak{o}(4;\mathbb C)$, in
particular for $\mathfrak{o}(4)$, $\mathfrak{o}(3,1)$ and $\mathfrak{o}(2,2)$, has been given in~\cite{BorowiecLukierskiTolstoy2016rmatrices,BorowiecLukierskiTolstoy2016rmatricesaddendum}.

We recall that the classification of  (A)dS DD $r$-matrices in~\cite{BHM2013cqg} was carried out in a basis with generators denoted  $\{J_0,J_1,J_2,P_0,P_1,P_2\}$. 
There are four DD structures for $\mathfrak{so}(3,1)$ (cases A, B, C and D), and three for AdS
 $\mathfrak{so}(2,2)$ (cases E, F and G).  Depending on the case considered, the relationship of this basis with the kinematical one used throughout the present paper is established by means of the  following   isomorphisms:
\be
\mbox{Cases A, C, E, F, G:}\quad
J_0 \rightarrow J, \quad J_1 \rightarrow -K_2,\quad J_2 \rightarrow K_1, \quad P_i \rightarrow P_i, \quad i=0,1,2 .
\label{csiso}
\ee
\be
\mbox{Cases B, D:}\quad J_0 \rightarrow J, \quad J_1 \rightarrow \frac 1{\eta} P_2,\quad J_2 \rightarrow - \frac 1{\eta} P_1, \quad P_0\rightarrow -P_0, \quad P_1\rightarrow \eta K_1, \quad P_2\rightarrow \eta K_2 , \quad \eta=\sqrt{\Lambda} .
\label{csiso2}
\ee
By applying these transformations onto the brackets (2.3) in~\cite{BHM2013cqg}, we  find that the commutation relations for the (2+1) (A)dS Lie algebras adopt the form
\begin{align} 
\begin{array}{lll}
 [J,K_1]=K_2, & \quad [J,K_2]=-K_1,  & \quad  [K_1, K_2]=-J,  \\[2pt]
 [J,P_0]=0 ,& \quad [J,P_1]=P_2 ,& \quad [J, P_2]=-P_1 ,\\[2pt]
[K_1,P_0]=P_1 ,& \quad [K_1,P_1]=P_0 ,& \quad [K_1, P_2]=0,\\[2pt]
[K_2,P_0]=P_2 ,& \quad[K_2,P_1]= 0 ,& \quad[K_2, P_2]=P_0,\\[2pt]
[P_0,P_1]=-\Lambda K_1, & \quad[P_0, P_2]=-\Lambda K_2 ,& \quad  [P_1,P_2]=\Lambda J .
 \end{array}
\label{ads}
\end{align}
Therefore, when $\Lambda<0$ we recover the AdS Lie algebra $\mathfrak{so}(2,2)$, for $\Lambda>0$ the dS   algebra $\mathfrak{so}(3,1)$, and the contraction $\c\to 0$ gives the Poincar\'e Lie algebra in a basis which is just~(\ref{aa}). Now, by using (\ref{csiso}) and (\ref{csiso2}), we rewrite in the kinematical basis (\ref{ads}) all the DD (A)dS $r$-matrices obtained in~\cite{BHM2013cqg}, namely
\begin{align}
\begin{array}{ll}
{\rm dS}\equiv \mathfrak{so}(3,1) \  \ (\c>0):  & r_{\rm A}'= \sqrt{\Lambda}\, K_1 \wedge K_2 + \frac12  ( -P_0 \wedge J - P_1 \wedge K_2 + P_2 \wedge K_1 ),\\[4pt]
 &r_{\rm B}'= \frac1 {\sqrt{\Lambda}} \, P_2\wedge P_1 +\frac 12  (- P_0 \wedge J + P_1 \wedge K_2 - P_2 \wedge K_1  ),\\[4pt]
 &r_{\rm C}'= \frac12  ( P_0 \wedge K_2 + P_1 \wedge J - P_2 \wedge K_1  ),\\[4pt]
 &r_{\rm D}'= {\sqrt{\Lambda}}\, J\wedge K_1+\frac 1  {\sqrt{\Lambda}} \, P_2\wedge P_0+\frac{(1+\mu^2)}{2\mu}\,  P_1 \wedge K_2\\[4pt]
   &\qquad\qquad +\frac{(\mu^2-1)}{2\mu}\,(- P_2\wedge J+P_0\wedge K_1),\quad \mu>0 .\\[6pt]
{\rm AdS}\equiv \mathfrak{so}(2,2) \  \ (\c<0):    &r_{\rm E}'= \sqrt{-\Lambda}\, J \wedge K_1 + \frac12  ( -P_0 \wedge J - P_1 \wedge K_2 + P_2 \wedge K_1 ),\\[4pt]
 &r_{\rm F}'= \frac12  ( P_0 \wedge K_2 + P_1 \wedge J - P_2 \wedge K_1  ) ,\\[4pt]
  &r_{\rm G}'=\frac{(1+\rho^2)}{4}\, (P_0\wedge K_2 + P_1\wedge J)-\frac \rho 2\, P_2  \wedge K_1 \\[4pt]
    &\qquad\qquad +\frac{(1-\rho^2)}{4\sqrt{-\Lambda}} \, (\Lambda\, J\wedge K_2 +P_0 \wedge P_1),\quad
    -1<\rho<1 .
 \end{array}
 \label{eq:rAdS}
\end{align}

Now let us analyse the vanishing cosmological constant limit $\Lambda\to0$ of all these expressions. Firstly, we obtain that
$$
 \lim_{\Lambda\to 0} r_{\rm A}' = \lim_{\Lambda\to 0} r_{\rm E}'=\tfrac12  ( -P_0 \wedge J - P_1 \wedge K_2 + P_2 \wedge K_1 ) \equiv  r_{0}' ,
$$
 which is just the Poincar\'e $r$-matrix (\ref{twists}) of Case 0 coming from the DD of $\mathfrak{sl}(2,\mathbb{R})$ and trivial cocommutator. Secondly,  we have that
$$
 \lim_{\Lambda\to 0} r_{\rm C}' = \lim_{\Lambda\to 0} r_{\rm F}'= \tfrac12  ( P_0 \wedge K_2 + P_1 \wedge J - P_2 \wedge K_1  ) \propto  r_{2}' ,
 \label{CF}
$$
which thus corresponds to Case 2 with the $r$-matrix expressed in the form (\ref{r2skew2}). This, in turn, means that 
we have obtained a common DD $r$-matrix for the   (A)dS and Poincar\'e Lie algebras that is just   a 
twisted version of the `space-like'  $\kappa$-(A)dS and  $\kappa$-Poincar\'e $r$-matrices studied in~\cite{BHMN2014sigma}. 

Finally, cases B, D and G seem to give rise to divergencies in the limit  $\c\to 0$. However we can rescale globally these $r$-matrices (the multiplication of a given $r$-matrix by a constant is also an $r$-matrix) in such a way that the limit does exist, namely
$$
 \lim_{\Lambda\to 0} \sqrt{\c}\,r_{\rm B}' =P_2\wedge P_1 ,\qquad
 \lim_{\Lambda\to 0} \sqrt{\c}\,r_{\rm D}' =P_2\wedge P_0 ,\qquad \lim_{\Lambda\to 0} \sqrt{-\c}\, r_{\rm G}' =  \frac{(1-\rho^2)}{4 } \,  P_0 \wedge P_1 .
$$
None of these limits provides a Poincar\'e $r$-matrix coming from a DD structure, and all of them belong to Class (V) in~\cite{Stachura1998} (see (\ref{classV})).

%%%%%%%%%%%%%%%%%%%%%%%%%%%%%%%%%%%%%%%%%%%%%%%

\sect{The   extended (1+1) Poincar\'e  algebra as a Drinfel'd double} \label{DD1+1}

Since the (1+1) Poincar\'e algebra 
\be
[K,P_0] = P_1, \qquad [K,P_1] = P_0, \qquad [P_1,P_0] = 0,  
\ee
is odd-dimensional,   no DD structure~\eqref{agd} can be defined within it. Nevertheless, if we consider the non-trivial central extension of the Poincar\'e Lie algebra $\overline{\mathfrak{iso}(1,1)}=\overline{\mathfrak{p}(1+1)}$ given by 
\be
[K,P_0] = P_1, \qquad [K,P_1] = P_0, \qquad [P_1,P_0] = F,  \qquad [F,\cdot] = 0,
\label{nw}
\ee
we will show in the sequel that the new central generator $F$ allows the introduction of two non-equivalent DD structures.
This extended algebra is also called the Nappi-Witten Lie algebra \cite{NappiWitten1993} and plays a relevant role in (1+1) gravity \cite{CJ1993extendedpoincare}. Casimir operators for the algebra~\eqref{nw} are given by: 
\be
C_1= P_0^2-P_1^2+F\,K+K\,F,
\qquad\qquad
C_2=F,
\label{extcas}
\ee
and the first of them already suggests the existence of a nondegenerate symmetric bilinear form underlying possible DD structures. 

The corresponding (1+1)-dimensional Poincar\'e group with a nontrivial central extension, $\overline{ISO(1,1)}= \overline{P(1+1)}$, is obtained by considering the  faithful representation $\rho:\overline{\mathfrak{p}(1+1)} \rightarrow \text{End} (\mathbb{R}^4)$ given by
\begin{align}
& \rho (F)=
\begin{pmatrix}
    0 & 0 & 0 & 0 \\
    0 & 0 & 0 & 0 \\
    0 & 0 & 0 & 0 \\
    -2 & 0 & 0 & 0 
  \end{pmatrix}, \qquad \rho (K)=
  \begin{pmatrix}
    0 & 0 & 0 & 0 \\
    0 & 0 & 1 & 0 \\
    0 & 1 & 0 & 0 \\
    0 & 0 & 0 & 0
  \end{pmatrix}
 , \nonumber \\[4pt]
&\rho (P_0)=
  \begin{pmatrix}
    0 & 0 & 0 & 0 \\
    1 & 0 & 0 & 0 \\
    0 & 0 & 0 & 0 \\
    0 & 0 & 1 & 0
  \end{pmatrix}, \qquad\ 
\rho (P_1)=
  \begin{pmatrix}
    0 & 0 & 0 & 0 \\
    0 & 0 & 0 & 0 \\
    1 & 0 & 0 & 0 \\
    0 & -1 & 0 & 0
  \end{pmatrix} ,
\end{align}
along with local coordinates on the Lie group $\{\phi,\xi,x^0,x^1\}$ associated with the generators $\{F,K,P_0,P_1\}$, respectively.  Hence we obtain the group element
\be
G=\exp \left( \phi \, \rho (F) \right) \exp \left( x^0 \, \rho (P_0) \right) \exp \left( x^1 \,  \rho (P_1) \right) \exp \left( \xi \, \rho (K) \right) ,
\ee
namely,
\be
G=
  \begin{pmatrix}
    1 & 0 & 0 & 0 \\
    x^0 & \cosh \xi & \sinh \xi & 0 \\
    x^1 & \sinh \xi & \cosh \xi & 0 \\
    x^0 \, x^1 - 2 \phi & -x^1 \cosh \xi + x^0 \sinh \xi & x^0 \cosh \xi - x^1 \sinh \xi & 1
  \end{pmatrix} .
\ee
From it, left- and right-invariant vector fields on the $\overline{P(1+1)}$ group are found to be
\begin{align}
\label{eqn:leftinvatiantpoincare1+1nontriviallyextended}
\begin{split}
\nabla^L_F &= \partial_\phi , \qquad \nabla^L_K = \partial_\xi  ,
\\
\nabla^L_{P_0} &= \cosh \xi \left( x^1 \, \partial_\phi + \partial_{x^0} \right) + \sinh \xi \, \partial_{x^1} ,
\\
\nabla^L_{P_1} &= \sinh \xi \left( x^1 \, \partial_\phi + \partial_{x^0} \right) + \cosh \xi \, \partial_{x^1}  ,
\end{split}
\end{align} 
\begin{align}
\label{eqn:rightinvatiantpoincare1+1nontriviallyextended}
   \nabla^R_K &= \tfrac12 \bigl( (x^0)^2 + (x^1)^2 \bigr) \partial_\phi +x^1  \partial_{x^0} +x^0  \partial_{x^1} + \partial_\xi ,
\nonumber \\
 \nabla^R_F &= \partial_\phi , \qquad  \nabla^R_{P_0} = \partial_{x^0} , \qquad  
\nabla^R_{P_1} = x^0 \partial_\phi + \partial_{x^1} .
\end{align}

In this context, we   define the  (1+1) Minkowski spacetime ${M}^{1+1}$ and its extended counterpart  $\overline{M}^{1+1}$ (see~\cite{BM2018extended}) as the following 
quotients by the Lorentz subalgebra $\mathfrak {h}$ and by the trivially extended  one $ \mathfrak{\overline h}$:
\begin{align}
{M}^{1+1}& = \overline{P(1+1)} /  { \overline H} ,&& \mathfrak{\overline h}= {\rm Lie}(\overline H)=\mathfrak{so} (1,1 )\oplus \mathbb R=\spn\{ K,F \},&& \mbox{coordinates:}\  x^0,x^1 . \nonumber \\
\overline{M}^{1+1}& = \overline{P(1+1)} / H, &&\mathfrak{ h}= {\rm Lie}( H)=\mathfrak{so} (1,1 )=\spn\{ K \},&& \mbox{coordinates:}\ x^0,x^1,\phi  .\label{mspace}
\end{align}

%%%%%%%%%%%%%%%%%%%%%%%%%%%%%%%%%%%%%%%%%%%%%%%

\subsection{Two-dimensional real Lie bialgebras and their Drinfel'd double structures}

The only two-dimensional non-Abelian real Lie algebra is the so-caled $\mathfrak{b}_2$ algebra with bracket
\be
[Y_1,Y_2]= Y_2.
\ee
It is also known that there exists, up to isomorphism, three real Lie bialgebra structures $\delta$ for this algebra, which are the `trivial one' with $\delta(Y_i)=0$ plus the two non-trivial ones given in~\cite{Gomez2000}. 
As we will show in what follows, two of these Lie bialgebras have the centrally extended (1+1)  Poincar\'e algebra~\eqref{nw} as its DD algebra. Explictly, with the notation used in~\cite{Gomez2000} these two Lie bialgebras are:
\begin{itemize}

\item Case 0: $\left( \mathfrak{g},\mathfrak{g^*} \right) = \left( \mathfrak{b}_2, \mathbb R^2\right)$ with $\chi = 0$, $\tilde{\delta} = 0$.

\item Case 1: $\left( \mathfrak{g},\mathfrak{g^*} \right) = \left( \mathfrak{b}_2,\mathfrak{b}_2\right)$ with $ \chi = e_0 \wedge e_1$, $\tilde{\delta} = 0$.

\end{itemize}
The remaining two-dimensional real Lie bialgebra structure was shown in~\cite{BCO2005quantizationDD} to have as its DD Lie algebra a central extension of $\mathfrak{sl}(2,\mathbb R)$, which is just the centrally extended (1+1) (A)dS Lie algebra. 
Note that the classification of nonisomorphic four-dimensional real DD structures was also given in~\cite{HS2002tdual2D}.

Commutation rules for these two DD   structures for   $\overline{\mathfrak{p}(1+1)}$ are given in Table \ref{table:DD1+1}, and in the following we will present these two structures in the kinematical basis, together with the associated classical $r$-matrices and (extended) noncommutative Minkowski spacetimes. We recall that the extended noncommutative Minkowski spacetimes studied in~\cite{BM2018extended} come from a (different) trivial central extension of the (1+1) Poincar\'e group.

%%%%%%%%%%%%%%%%%%%%%%%%%%%%%%%%%%%%%%%%%%%%%%%

\begin{table}[tp]{\footnotesize
\caption{\small The two non-equivalent DD Lie algebras which are isomorphic to the extended (1+1) Poincar\'e algebra.} \label{table:DD1+1}
\begin{center}
\begin{tabular}{lll}
\hline 
 &  &    \\[-1.5ex]
 & Case 0 & Case 1 \\[3pt]
\hline
 &  &    \\[-1.5ex] 
$[Y_1,Y_2]$ & $Y_2$ & $Y_2$ \\[3pt] 
\hline 
 &  &    \\[-1.5ex]
$[y^1,y^2]$ & $0$ & $y^1$  \\[3pt]
\hline 
 &  &   \\[-1.5ex]
$[y^1,Y_1]$ & $0$ & $-Y_2$   \\[3pt]
$[y^1,Y_2]$ & $0$  & $0$\\[3pt] 
$[y^2,Y_1]$ & $y^2$ & $y^2+Y_1$  \\[3pt]
$[y^2,Y_2]$ & $-y^1$ & $-y^1$  \\[3pt] 
\hline 
\end{tabular} 
\end{center}}
\end{table}

%%%%%%%%%%%%%%%%%%%%%%%%%%%%%%%%%%%%%%%%%%%%%%%

\subsection{Case 0}

A Lie algebra isomorphism is given by 
\begin{align}
&K= - Y_1, & 
&P_0=\frac{1}{\sqrt{2}} (y^2 + Y_2), & 
&P_1=\frac{1}{\sqrt{2}} (y^2 - Y_2), & 
&F=- y^1,
\end{align}
that from (\ref{ages})  gives the following non-zero entries for the pairing
\begin{align}
\label{pairing1+1}
&\langle K,F\rangle =1, & 
&\langle P_0,P_0\rangle =1, & 
&\langle P_1,P_1\rangle =-1.
\end{align}
By   inserting the inverse isomorphism   in~\eqref{rcanon} we obtain the classical $r$-matrix
\be
r_0=\sum_{i=1}^2{y^i\otimes Y_i}= K \otimes F + \frac12 \left(P_0 \otimes P_0 - P_1 \otimes P_1 + P_0 \wedge P_1 \right),
\label{r0_1+1}
\ee
and by subtracting the tensorized Casimirs $C_1$ (\ref{extcas})    we obtain the skew-symmetric $r$-matrix
\be
r'_0=\frac12 (K \wedge F + P_0 \wedge P_1) .
\label{r0skew_1+1}
\ee
The DD cocommutator reads
\begin{align}
\label{eqn:cocom_poincare1+1nontriviallyextended_Case0}
\begin{split}
\delta_D (K)&=0, \qquad \delta_D (F)=0,\\
\delta_D (P_0)&= - \tfrac12 \left( P_0 \wedge F + P_1 \wedge F \right), \\
\delta_D (P_1)&= - \tfrac12 \left( P_0 \wedge F + P_1 \wedge F \right) .
\end{split}
\end{align}
Since $\delta_D (K)=\delta_D (F)=0$, we have that, trivially, $\delta_D (\mathfrak{\overline h}) \subset \mathfrak {\overline h} \wedge \mathfrak{\overline h}$ and the associated Poisson Minkowski spacetime is a Poisson subgroup one. Obviously,   $\delta_D (\mathfrak h) \subset \mathfrak h \wedge \mathfrak h$ and the  Poisson extended Minkowski spacetime is of Poisson subgroup type as well. 

A two-parameter generalization  of the $r$-matrix  \eqref{r0skew_1+1} fulfilling the modified CYBE is given by
\be
r_{\rm 0, (\pa_0,\pb_0)} ' =\pa_0 K\wedge F + \pb_0 P_0 \wedge P_1 ,
\ee
with $\pa_0,\pb_0 \in \mathbb R$. The associated fundamental Poisson brackets for the   coordinates $\{x^0,x^1,\phi\}$ (see (\ref{mspace}))  are obtained from the  Sklyanin bracket  and 
turn out to be
\be
\{x^0,x^1 \} = 0,\qquad \{\phi,x^0\}=\pb_0 x^0+\pa_0 x^1,\qquad \{\phi,x^1\}=\pa_0 x^0+\pb_0 x^1,
\label{ecase0}
\ee
which are linear and can be straightforwardly quantized. Hence, although the Minkowski coordinates $\{x^0,x^1\}$ Poisson-commute, the  extended Minkowski spacetime $\{x^0,x^1,\phi\}$ defines a noncommutative structure, thus suggesting that  the role of central extensions in noncommutative spacetimes deserves a deeper analysis along the lines presented in~\cite{BM2018extended}. Notice also that for the DD spacetime, obtained when $\pa_0=\pb_0$, the Poisson algebra (\ref{ecase0}) is isomorphic to $\mathfrak{b}_2\oplus \mathbb R$ such that
$\mathfrak{b}_2=\spn\{ \phi,  x^0+ x^1\} $ and $\mathbb R=\spn\{ x^0- x^1 \}$.

%%%%%%%%%%%%%%%%%%%%%%%%%%%%%%%%%%%%%%%%%%%%%%%

\subsection{Case 1}

The Lie algebra isomorphism is now given by 
\begin{align}
&K= - Y_1, & 
&P_0= \frac{1}{\sqrt 2} (y^2 + Y_1+Y_2), & 
&P_1=\frac{1}{\sqrt 2} (y^2+Y_1-Y_2), & 
&F= -y^1 + Y_2,
\end{align}
and the pairing is exactly \eqref{pairing1+1}.

The inverse of the above isomorphism   
 inserted into~\eqref{rcanon} gives rise to the classical $r$-matrix 
\be
r_1= F \otimes K + \frac{1}{\sqrt 2} \left( K \wedge P_0 + P_1 \wedge K \right) + \frac12 \left( P_1 \wedge P_0 + P_0\otimes P_0 - P_1 \otimes P_1 \right),
\label{r1_1+1}
\ee
and by subtracting the tensorized Casimir $C_1$    (\ref{extcas})   we obtain the skew-symmetric $r$-matrix
\be
r'_1=\frac12 (F \wedge K + P_1 \wedge P_0) + \frac{1}{\sqrt 2} (P_1 \wedge K+K \wedge P_0).
\label{r1skew_1+1}
\ee
The DD cocommutator is
\begin{align}
\label{eqn:cocom_poincare1+1nontriviallyextended_Case1}
\begin{split}
\delta_D (K)&=\frac{1}{\sqrt 2} \left( -K \wedge P_0 + K \wedge P_1 \right), \qquad \delta_D (F)=0 ,\\
\delta_D (P_0)&=\frac{1}{\sqrt 2} \left( P_0 \wedge P_1 + K \wedge F \right)+\frac12 \left( P_0\wedge F + P_1 \wedge F \right), \\
\delta_D (P_1)&=\frac{1}{\sqrt 2} \left( P_0 \wedge P_1 + K \wedge F \right)+\frac12 \left( P_0 \wedge F + P_1 \wedge F \right), 
\end{split}
\end{align}
which is of `true' coisotropic type for both Minkowski and extended Minkowski spacetimes (\ref{mspace}).

The $r$-matrix~\eqref{r1skew_1+1} can be generalized to the following two-parameter solution of the modified CYBE
\be
r'_{\rm 1, (\pa_1,\pb_1)} =\pa_1 (K\wedge F + P_0 \wedge P_1) + \pb_1 (K\wedge P_1+P_0 \wedge K) ,
\label{er1}
\ee
with $\pa_1,\pb_1\in \mathbb R$ (the DD case corresponds to set $\pb_1=\sqrt{2}\pa_1$). The associated fundamental Poisson brackets for the   coordinates $x^0,x^1,\phi $ are given by
\begin{align}
\{x^0,x^1 \} &= -\pb_1 (x^0 + x^1),\nonumber\\
\{\phi,x^0 \} &= \pa_1 (x^0+x^1) +\tfrac 12 \pb_1 \bigl(x^0+x^1\bigr)^2, \nonumber\\
\{\phi,x^1 \} &= \pa_1 (x^0+x^1) +\tfrac 12 \pb_1  (x^0+x^1 ) (x^0-x^1 ) .
\label{case1pb}
\end{align}
Therefore the first bracket defines the Poisson Minkowski spacetime and  can be trivially quantized.  We remark that  this comes from the $\pb_1$-term in (\ref{er1}), which is a solution of the CYBE, that corresponds to the so-called `null-plane' noncommutative Minkowski spacetime studied in \cite{BHOPS1995tmatrix} (see~\cite{BHP1997nullplane} for its (3+1) generalization). In light-cone coordinates $x^\pm=x^0\pm x^1$ the Poisson structure takes the simpler form
\be
\{x^+,x^- \} =2\pb_1 x^+ .
\ee

Interestingly enough, the Poisson extended Minkowski spacetime defined by the three brackets (\ref{case1pb}) is quadratic and cannot  be straightfordwardly quantized. Notice that  the $\pa_1$-term in the $r$-matrix (\ref{er1})  is just the one for the above  Case 0  (\ref{r0skew_1+1}) and, consequently,~\eqref{er1}  can be regarded either as a generalization of Case 0 with additional deformation parameter $\pb_1$, or as a generalized `null-plane' Minkowski spacetime with additional parameter $\pa_1$.

%%%%%%%%%%%%%%%%%%%%%%%%%%%%%%%%%%%%%%%%%%%%%%%

\section{Concluding remarks} \label{concluding}

In this paper the complete set of DD structures for the Poincar\'e group  in (2+1) dimensions and for its non-trivial central extension in (1+1) dimensions have been presented. For every nonisomorphic DD structure we have worked out an explicit isomorphism between the canonical basis in the double and the Poincar\'e kinematical basis. By making use of the classical $r$-matrices arising from the DD structures that fulfil the coisotropy condition with respect to the Lorentz subalgebra, we have constructed the associated DD Poisson Minkowski spacetimes. These results complete the chart of DD structures for the Poincar\'e group, since they do not exist in $(N+1)$ dimensions with $N\geq 3$.

In the (2+1)-dimensional case, DD structures have a remarkable interest due to their connection with the Chern-Simons approach to gravity in (2+1) dimensions, in which the gauge group is identified with the isometry group. Indeed, it is the existence of a DD structure that guarantees the Fock-Rosly conditions for the $r$-matrix defining the Poisson structure on the phase space of the theory. We have explicitly constructed the eight nonisomorphic DD structures on $\mathfrak{iso}(2,1)$, displayed in Table~\ref{table:2+1_r-matrices}, meanwhile in~\cite{BHM2013cqg} the four nonisomorphic DD structures for $\mathfrak{so}(3,1)$ and the three ones for $\mathfrak{so}(2,2)$ were deduced. Therefore, this paper completes the study of DD structures on the three Lorentzian kinematical groups in (2+1) dimensions, and shows how to connect Poincar\'e DD structures with (A)dS ones through the contraction induced by the vanishing cosmological constant limit $\c\to 0$. In particular, it is found that only two of the eight Poincar\'e DD $r$-matrices (Cases 0 and 2)  can be obtained through such a contraction procedure.

For each of the eight different DD structures, we have also  investigated whether the corresponding Lie bialgebra was coisotropic with respect to the Lorentz subalgebra, i.e.~if the condition $\delta_D (\mathfrak{h}) \subset \mathfrak{h} \wedge \mathfrak{g}$ holds. The result is that five out of the eight DD structures do fulfil this condition, as shown in Table~\ref{table:2+1_r-matrices}, thus providing five (noncommutative) Poisson Minkowski  spacetimes. Only two of them (Cases 0 and   1) fulfil the stronger requirement of being generated by a Lorentz Poisson subgroup, i.e $\delta_D (\mathfrak{h}) \subset \mathfrak{h} \wedge \mathfrak{h}$. Case 0 is the DD obtained from a $\mathfrak{sl}(2,\mathbb{R})$ Lie bialgebra with a trivial cocommutator map, and was previously known in the (2+1) quantum gravity literature. Interestingly enough, Case 1 gives rise to a quadratic Poisson Minkowski spacetime which, to the best of our knowledge, has not been considered previously in the literature and whose quantization deserves further study. The remaining three coisotropic DD structures (Cases 2, 6 and 7) provide noncommutative spacetimes of Lie-algebraic type, whose quantization is straightforward.

  In   (1+1) dimensions, only three nonisomorphic DD structures do exist and two of them correspond to the centrally extended (1+1) Poincar\'e Lie algebra $\overline{\mathfrak{p}(1+1)}$. The remaining four-dimensional DD Lie algebra leads to $\overline{\mathfrak{so}(2,1)}\simeq \mathfrak{gl}(2,\mathbb R)$ (see~\cite{BCO2005quantizationDD,HS2002tdual2D}), which is the centrally extended Lie algebra of isometries of (1+1) (A)dS spacetimes. For the two Poincar\'e structures the coisotropy condition holds and (extended) Poisson Minkowski spacetimes can be explicitly constructed. Moreover, in Case 0 the noncommutative Minkowski space  is  generated by a commutative  Poisson subgroup, while the extended space is of Lie-algebraic type. Case 1 gives rise to the `null-plane' noncommutative Minkowski spacetime, and its extended version is again defined by a quadratic Poisson algebra. Therefore, the study of DD structures for (1+1) Lorentzian groups has been also completed.

It is worth mentioning that the existence of DD structures for kinematical groups is ruled by the existence of a nondegenerate, symmetric, and `associative' --in the sense of \eqref{associative}-- bilinear form. We stress that, in contradistinction with the rich variety of DD structures on the (2+1) Poincar\'e algebra, it is easy to see that in the (3+1) case no DD structures can be found, since no such bilinear form exists for $\mathfrak{iso}(3,1)$ (see~\cite{Figueroa-OFarrill2018, AF2018}). Furthermore, neither the static, nor the Galilean and Newton-Hooke kinematical algebras admit any DD structure in (2+1) and (3+1) dimensions,  and that neither Carroll nor the Euclidean Lie algebras admit DD structures in (3+1) dimensions. Thus, we have that the Galilean limit of all the DD $r$-matrices for $\mathfrak{iso}(2,1)$ obtained in this paper would lead (in case that such a limit does converge) to Galilean $r$-matrices which would not come from a DD structure. 
In this respect, we also point out that DD structures for the twice extended (2+1) Galilei algebra do exist, and one of them was fully constructed in~\cite{BHN2014cqg}, thus providing a meaningful connection with previous quantum group models for Galilean (2+1) gravity~\cite{PS2009approach,PS2010}. Indeed, it could happen that this `exotic' Galilean DD could be obtained as the appropriate contraction of a relativistic DD structure for a twice (trivially) extended (2+1) Poincar\'e algebra, a problem that would also deserve some attention in the future.

Moreover, although no DD structure exists for $\mathfrak{iso}(3,1)$, the problem of the classification of DD structures on both $\mathfrak{so}(4,1)$ and $\mathfrak{so}(3,2)$ is worth to be faced, and should be based in the classification of Lie bialgebra structures for 5-dimensional real Lie algebras. We recall that a first DD structure on $\mathfrak{so}(3,2)$ was fully worked out in the kinematical basis in \cite{BHN2015towards31,BHMN2017kappa3+1}. In fact, as proved in \cite{Figueroa-OFarrill2018higher}, for kinematical Lie algebras in (N+1) dimensions ($N \ge 4$), only for $\mathfrak{so}(N+1,1)$, $\mathfrak{so}(N+2)$ and $\mathfrak{so}(N,2)$ such a suitable nondegenerate  bilinear form does exist. Therefore these three Lie algebras could admit DD structures in higher dimensions.

Two more lines of research are suggested by the results here presented. The first of them arises from the fact that all the DD $r$-matrices that we have obtained are natural candidates for generating quantum Poincar\'e deformations which are quasitriangular Hopf algebras endowed with a quantum universal $R$-matrix. In this respect, the underlying DD structure should be helpful in order to construct such quantum Poincar\'e algebras and their associated quantum $R$-matrices. Note that both Cases 0 and 1 in (2+1) dimensions turn out to be particularly interesting, since for both of them the Lorentz subgroup would be promoted to a quantum subgroup once the full quantum Poincar\'e Hopf algebra had been constructed.

The second one points towards the well-known connection between Manin triples and the non-Abelian version of Poisson-Lie $T$-dual $\sigma$-models (see~\cite{HS2002tdual2D,KS1995duality,Sfetsos1998,LledoVaradarajan1998,JafarizadehRezaei1999,BeggsMajid2001tduality,HS2004tplurality3D} and references therein). We recall that these $\sigma$-models for four-dimensional DDs were constructed in~\cite{HS2002tdual2D}, and among them we can find the two ones corresponding to our Cases 0 and  1 for the centrally extended (1+1) Poincar\'e group. In the same manner, $T$-dual $\sigma$-models having the (2+1) Poincar\'e group as their DD Lie group would be the ones that come from the eight DD structures presented in this paper. This `eightfold' non-Abelian Poisson-Lie $T$-plurality   would be worth studying, since --to the best of our knowledge-- none of these eight $\sigma$-models with (2+1) Poincar\'e DD symmetry has been previously constructed in the literature.

%%%%%%%%%%%%%%%%%%%%%%%%%%%%%%%%%%%%%%%%%%%%%%%

\section*{Acknowledgements}

This work was partially supported by Ministerio de Ciencia, Innovaci\'on y Universidades (Spain) under grant MTM2016-79639-P (AEI/FEDER, UE), by Junta de Castilla y Le\'on (Spain) under grant BU229P18 and by the Action MP1405 QSPACE from the European Cooperation in Science and Technology (COST). I.G-S. acknowledges a predoctoral grant from Junta de Castilla y Le\'on and the European Social Fund. The authors are grateful to M. Bordemann, B.J. Schroers and J. Lukierski for helpful discussions.

%%%%%%%%%%%%%%%%%%%%%%%%%%%%%%%%%%%%%%%%%%%%%%%

\section*{Appendix. Stachura classification of (2+1) Poincar\'e $r$-matrices}

\setcounter{equation}{0}
\renewcommand{\theequation}{A.\arabic{equation}}

In the paper~\cite{Stachura1998} the complete classification of equivalence classes (under automorphisms) of skew-symmetric $r$-matrices for the (2+1) Poincar\'e algebra was given. In order to translate this classification in terms of the kinematical basis we have used throughout the paper, the appropriate isomorphism is given by the map 
\be
e_1 = P_0, \qquad e_2 = P_1, \qquad e_3 = P_2, \qquad
k_1 = -J, \qquad k_2 = -K_2, \qquad k_3 = K_1,
\label{changeS}
\ee
with  $\{e_1,e_2,e_3, k_1,k_2,k_3\}$ being  the basis used in~\cite{Stachura1998}.
The classification is based in expressing Poincar\'e $r$-matrices as the sum of three contributions
\be
r=a+b+c,
\qquad
\qquad
a\subset \mathfrak t \wedge \mathfrak t,
\qquad
b\subset \mathfrak t \wedge \mathfrak h,
\qquad
c\subset \mathfrak h \wedge \mathfrak h,
\ee
where $\mathfrak t =\spn\{P_0,P_1,P_2 \}$ and $\mathfrak h =\spn\{ J,K_1,K_2\}$.
It turns out that the classification is ruled by two real parameters $\{\mu,p\}$ defined by 
\be
[[a,b]]= p \,\tilde \eta, \qquad 2 [[a,c]] + [[b,b]] = \mu\, \Omega, \qquad \tilde \eta \in \bigwedge^3 \mathfrak{t}, \qquad \Omega \in \bigwedge^2 \mathfrak{t} \otimes \mathfrak{h},
\ee
and the cases with $p=0$ correspond to $r$-matrices giving rise to coisotropic Lie bialgebras with respect to $\mathfrak h$.

The explicit expressions for the eight equivalence classes of (2+1) Poincar\'e $r$-matrices are the following: 

\noindent{\bf Class (I)}:
\be
c=\tfrac{1}{\sqrt 2}\, K_1 \wedge (J+K_2), 
\qquad
b=\alpha (-P_0 \wedge J - P_1 \wedge K_2 + P_2 \wedge K_1), 
\qquad a=0.
\label{classI}
\ee
Here $\alpha=\{0,1\}$, $\mu=2\alpha^2$, $p=0$. This is the only case in the classification with $c\neq 0$, so hereafter $c=0$.

 \medskip
 
\noindent{\bf Class (IIa)}:
\be
b=\rho P_2 \wedge K_1 - \alpha (P_1 \wedge J + P_0 \wedge K_2), 
\qquad
a=a_{01} P_0\wedge P_1 + a_{02} P_0\wedge P_2 + a_{12} P_1\wedge P_2 ,
\label{classIIa}
\ee
with $\alpha=\{0,1\}$, $\rho \ge 0$, $\alpha^2+\rho^2\ne 0$, $\mu=-2\alpha^2$, $p\in\mathbb{R}$,  and where from now on  $\{a_{01},a_{02},a_{12}\}$ denote    free real parameters. Note that the automorphism  (\ref{auto2b}) transforms $\rho \ge 0$ into $\rho \le 0$. 
\medskip

\noindent{\bf Class (IIb)}:
\be
b=-\rho P_0 \wedge J - \alpha (P_1 \wedge K_1+P_2 \wedge K_2 ), 
\qquad
a=a_{01} P_0\wedge P_1 + a_{02} P_0\wedge P_2 + a_{12} P_1\wedge P_2,
\ee
with $\alpha=\{0,1\}$, $\rho \ge 0$, $\alpha^2+\rho^2\ne 0$,  $\mu=2\alpha^2$, $p\in \mathbb R$. 
 \medskip
 
 \noindent{\bf Class (IIc)}:
\be
b= \frac{\alpha}{\sqrt 2}\bigl(-P_2 \wedge (J+K_2)+ (P_0 - P_1) \wedge K_1\bigr)- \rho (P_0 -P_1) \wedge (J+K_2),
\qquad
a=a_{01} P_0\wedge P_1 + a_{02} P_0\wedge P_2 + a_{12} P_1\wedge P_2,
\ee
with $\alpha=\{0,1\}$, $\rho \ge 0$, $\alpha^2+\rho^2\ne 0$,   $\mu=0$, $p\in \mathbb R$. 
\medskip

\noindent{\bf Class (IIIa)}:
\be
b= \frac{1}{\sqrt 2} (P_0 - P_1) \wedge K_1,
\qquad
a=a_{01} P_0\wedge P_1 + a_{02} P_0\wedge P_2 + a_{12} P_1\wedge P_2,
\ee
with $\mu=0$, $p\in \mathbb R$. 
\medskip

\noindent{\bf Class (IIIb)}:
\be
b= -P_0 \wedge J - (\rho -1) P_1 \wedge J - (\rho +1) P_0 \wedge K_2 + P_1 \wedge K_2 + \rho P_2 \wedge K_1,
\qquad
a=a_{01} P_0\wedge P_1 + a_{02} P_0\wedge P_2 + a_{12} P_1\wedge P_2,
\label{classIIIb}
\ee
with $\rho \in \mathbb R^\ast$, $\mu=2\rho^2$, $p\in \mathbb R$. 
\medskip

\noindent{\bf Class (IV)}:
\be
b=-P_0 \wedge J - P_1 \wedge K_2 + P_2 \wedge K_1, 
\qquad
a = 0,
\label{classIV}
\ee
with $\mu=2$, $p=0$. 
\medskip

\noindent{\bf Class (V)}:
\be
b=0, 
\qquad
a=a_{01} P_0\wedge P_1 + a_{02} P_0\wedge P_2 + a_{12} P_1\wedge P_2,
\label{classV}
\ee
with $\mu=0$, $p=0$. 
 \medskip

As it is shown in Table~\ref{table:2+1_r-matrices}, we have proven that only four of the above classes contain DD structures. Among them, only Class (IV) is by itself a DD, while  
Classes (I), (IIa) and (IIIb)  contain DD structures  for some specific values of the parameters.

%%%%%%%%%%%%%%%%%%%%%%%%%%%%%%%%%%%%%%%%%%%%%%%


\begin{thebibliography}{10}

\bibitem{Drinfeld1987icm}
V.~Drinfeld.
\newblock {Quantum Groups}.
\newblock In {\em Proc. Int. Congr. Math. (Berkeley 1986)}, pages 798--820,
  Providence, 1987. American Mathematical Society.

\bibitem{TV1992invariants}
V.~G. Turaev and O.~Y. Viro.
\newblock {State sum invariants of 3-manifolds and quantum 6j-symbols}.
\newblock {\em Topology}, 31(4):865--902, 1992.
\newblock \href {http://dx.doi.org/10.1016/0040-9383(92)90015-A}
  {\path{doi:10.1016/0040-9383(92)90015-A}}.

\bibitem{AGS1995}
A.~Y. Alekseev, H.~Grosse, and V.~Schomerus.
\newblock {Combinatorial quantization of the Hamiltonian Chern-Simons theory
  I}.
\newblock {\em Commun. Math. Phys.}, 172(2):317--358, 1995.
\newblock \href {http://arxiv.org/abs/hep-th/9403066}
  {\path{arXiv:hep-th/9403066}}, \href {http://dx.doi.org/10.1007/BF02099431}
  {\path{doi:10.1007/BF02099431}}.

\bibitem{AGS1996}
A.~Y. Alekseev, H.~Grosse, and V.~Schomerus.
\newblock {Combinatorial quantization of the Hamiltonian Chern-Simons theory
  II}.
\newblock {\em Commun. Math. Phys.}, 174(3):561--604, 1996.
\newblock \href {http://arxiv.org/abs/hep-th/9408097}
  {\path{arXiv:hep-th/9408097}}, \href {http://dx.doi.org/10.1007/BF02101528}
  {\path{doi:10.1007/BF02101528}}.

\bibitem{AS1996duke}
A.~Y. Alekseev and V.~Schomerus.
\newblock {Representation theory of Chern-Simons observables}.
\newblock {\em Duke Math. J.}, 85(2):447--510, 1996.
\newblock \href {http://arxiv.org/abs/q-alg/9503016}
  {\path{arXiv:q-alg/9503016}}, \href
  {http://dx.doi.org/10.1215/S0012-7094-96-08519-1}
  {\path{doi:10.1215/S0012-7094-96-08519-1}}.

\bibitem{BR1995}
E.~Buffenoir and P.~Roche.
\newblock {Two dimensional lattice gauge theory based on a quantum group}.
\newblock {\em Commun. Math. Phys.}, 170(3):669--698, 1995.
\newblock \href {http://arxiv.org/abs/hep-th/9405126}
  {\path{arXiv:hep-th/9405126}}, \href {http://dx.doi.org/10.1007/BF02099153}
  {\path{doi:10.1007/BF02099153}}.

\bibitem{MS2003}
C.~Meusburger and B.~J. Schroers.
\newblock {The quantisation of poisson structures arising in Chern-Simons
  theory with gauge group $G\ltimes \mathfrak{g}^\ast$}.
\newblock {\em Adv. Theor. Math. Phys.}, 7(6):1003--1043, 2003.
\newblock \href {http://arxiv.org/abs/hep-th/0310218}
  {\path{arXiv:hep-th/0310218}}.

\bibitem{NR2003cosmological}
K.~Noui and P.~Roche.
\newblock {Cosmological deformation of Lorentzian spin foam models}.
\newblock {\em Class. Quantum Gravity}, 20(14):3175--3213, 2003.
\newblock \href {http://arxiv.org/abs/gr-qc/0211109}
  {\path{arXiv:gr-qc/0211109}}, \href
  {http://dx.doi.org/10.1088/0264-9381/20/14/318}
  {\path{doi:10.1088/0264-9381/20/14/318}}.

\bibitem{FM2012}
W.~J. Fairbairn and C.~Meusburger.
\newblock {Quantum deformation of two four-dimensional spin foam models}.
\newblock {\em J. Math. Phys.}, 53(2):022501, 2012.
\newblock \href {http://arxiv.org/abs/1012.4784} {\path{arXiv:1012.4784}},
  \href {http://dx.doi.org/10.1063/1.3675898} {\path{doi:10.1063/1.3675898}}.

\bibitem{Majid1988}
S.~Majid.
\newblock {Hopf algebras for physics at the Planck scale}.
\newblock {\em Class. Quantum Gravity}, 5(12):1587--1606, 1988.
\newblock \href {http://dx.doi.org/10.1088/0264-9381/5/12/010}
  {\path{doi:10.1088/0264-9381/5/12/010}}.

\bibitem{LRNT1991}
J.~Lukierski, H.~Ruegg, A.~Nowicki, and V.~N. Tolstoy.
\newblock {q-deformation of Poincar{\'{e}} algebra}.
\newblock {\em Phys. Lett. B}, 264(3-4):331--338, 1991.
\newblock \href {http://dx.doi.org/10.1016/0370-2693(91)90358-W}
  {\path{doi:10.1016/0370-2693(91)90358-W}}.

\bibitem{LNR1992fieldtheory}
J.~Lukierski, A.~Nowicki, and H.~Ruegg.
\newblock {New quantum Poincar{\'{e}} algebra and $\kappa$-deformed field
  theory}.
\newblock {\em Phys. Lett. B}, 293(3-4):344--352, 1992.
\newblock \href {http://dx.doi.org/10.1016/0370-2693(92)90894-A}
  {\path{doi:10.1016/0370-2693(92)90894-A}}.

\bibitem{BHOS1995nullplane}
A.~Ballesteros, F.~J. Herranz, M.~A. del Olmo, and M.~Santander.
\newblock {A new ``null-plane" quantum Poincar{\'{e}} algebra}.
\newblock {\em Phys. Lett. B}, 351(1-3):137--145, 1995.
\newblock \href {http://arxiv.org/abs/q-alg/9502019}
  {\path{arXiv:q-alg/9502019}}, \href
  {http://dx.doi.org/10.1016/0370-2693(95)00386-Y}
  {\path{doi:10.1016/0370-2693(95)00386-Y}}.

\bibitem{BRH2003minkowskian}
A.~Ballesteros, N.~R. Bruno, and F.~J. Herranz.
\newblock {A non-commutative Minkowskian spacetime from a quantum AdS algebra}.
\newblock {\em Phys. Lett. B}, 574(3-4):276--282, 2003.
\newblock \href {http://arxiv.org/abs/hep-th/0306089}
  {\path{arXiv:hep-th/0306089}}, \href
  {http://dx.doi.org/10.1016/j.physletb.2003.09.014}
  {\path{doi:10.1016/j.physletb.2003.09.014}}.

\bibitem{Amelino-Camelia2010symmetry}
G.~Amelino-Camelia.
\newblock {Doubly-special relativity: Facts, myths and some key open issues}.
\newblock {\em Symmetry}, 2(1):230--271, 2010.
\newblock \href {http://arxiv.org/abs/1003.3942} {\path{arXiv:1003.3942}},
  \href {http://dx.doi.org/10.3390/sym2010230} {\path{doi:10.3390/sym2010230}}.

\bibitem{LN2003versus}
J.~Lukierski and A.~Nowicki.
\newblock {Doubly special relativity versus $\kappa$-deformation of
  relativistic kinematics}.
\newblock {\em Int. J. Mod. Phys. A}, 18(1):7--18, 2003.
\newblock \href {http://arxiv.org/abs/hep-th/0203065v3}
  {\path{arXiv:hep-th/0203065v3}}, \href
  {http://dx.doi.org/10.1142/S0217751X03013600}
  {\path{doi:10.1142/S0217751X03013600}}.

\bibitem{FKS2004gravity}
L.~Freidel, J.~Kowalski-Glikman, and L.~Smolin.
\newblock 2+1 gravity and doubly special relativity.
\newblock {\em Phys. Rev. D}, 69(4):044001, 2004.
\newblock \href {http://arxiv.org/abs/hep-th/0307085}
  {\path{arXiv:hep-th/0307085}}, \href
  {http://dx.doi.org/10.1103/PhysRevD.69.044001}
  {\path{doi:10.1103/PhysRevD.69.044001}}.

\bibitem{Kowalski-Glikman2013living}
J.~Kowalski-Glikman.
\newblock {Living in curved momentum space}.
\newblock {\em Int. J. Mod. Phys. A}, 28(12):1330014, 2013.
\newblock \href {http://arxiv.org/abs/1303.0195} {\path{arXiv:1303.0195}},
  \href {http://dx.doi.org/10.1142/S0217751X13300147}
  {\path{doi:10.1142/S0217751X13300147}}.

\bibitem{GM2013relativekappa}
G.~Gubitosi and F.~Mercati.
\newblock {Relative locality in $\kappa$-Poincar{\'{e}}}.
\newblock {\em Class. Quantum Gravity}, 30(14):145002, 2013.
\newblock \href {http://arxiv.org/abs/1106.5710} {\path{arXiv:1106.5710}},
  \href {http://dx.doi.org/10.1088/0264-9381/30/14/145002}
  {\path{doi:10.1088/0264-9381/30/14/145002}}.

\bibitem{BGGH2017curvedplb}
A.~Ballesteros, G.~Gubitosi, I.~Guti{\'{e}}rrez-Sagredo, and F.~J. Herranz.
\newblock {Curved momentum spaces from quantum groups with cosmological
  constant}.
\newblock {\em Phys. Lett. B}, 773:47--53, 2017.
\newblock \href {http://arxiv.org/abs/1707.09600} {\path{arXiv:1707.09600}},
  \href {http://dx.doi.org/10.1016/j.physletb.2017.08.008}
  {\path{doi:10.1016/j.physletb.2017.08.008}}.

\bibitem{BGGH31}
A.~Ballesteros, G.~Gubitosi, I.~Guti{\'{e}}rrez-Sagredo, and F.~J. Herranz.
\newblock {Curved momentum spaces from quantum (Anti-)de Sitter groups in (3+1)
  dimensions}.
\newblock {\em Phys. Rev. D}, 97(10):106024, 2018.
\newblock \href {http://arxiv.org/abs/1711.05050} {\path{arXiv:1711.05050}},
  \href {http://dx.doi.org/10.1103/PhysRevD.97.106024}
  {\path{doi:10.1103/PhysRevD.97.106024}}.

\bibitem{DJH1984}
S.~Deser, R.~Jackiw, and G.~{'t Hooft}.
\newblock {Three-dimensional Einstein gravity: dynamics of flat space}.
\newblock {\em Ann. Phys. (N. Y).}, 152(1):220--235, 1984.
\newblock \href {http://dx.doi.org/10.1016/0003-4916(84)90085-X}
  {\path{doi:10.1016/0003-4916(84)90085-X}}.

\bibitem{Carlip2003book}
S.~Carlip.
\newblock {\em {Quantum gravity in 2+ 1 dimensions}}.
\newblock Cambridge University Press, 1998.
\newblock \href {http://dx.doi.org/10.1017/CBO9780511564192}
  {\path{doi:10.1017/CBO9780511564192}}.

\bibitem{AT1986}
A.~Ach{\'{u}}carro and P.~K. Townsend.
\newblock {A Chern-Simons action for three-dimensional anti-de Sitter
  supergravity theories}.
\newblock {\em Phys. Lett. B}, 180(1-2):89--92, 1986.
\newblock \href {http://dx.doi.org/10.1016/0370-2693(86)90140-1}
  {\path{doi:10.1016/0370-2693(86)90140-1}}.

\bibitem{Witten1988}
E.~Witten.
\newblock 2+1 dimensional gravity as an exactly soluble system.
\newblock {\em Nucl. Phys. B}, 311(1):46--78, 1988.
\newblock \href {http://dx.doi.org/10.1016/0550-3213(88)90143-5}
  {\path{doi:10.1016/0550-3213(88)90143-5}}.

\bibitem{AM1995moduli}
A.~Y. Alekseev and A.~Z. Malkin.
\newblock {Symplectic structure of the moduli space of flat connection on a
  Riemann surface}.
\newblock {\em Commun. Math. Phys.}, 169(1):99--119, 1995.
\newblock \href {http://arxiv.org/abs/hep-th/9312004}
  {\path{arXiv:hep-th/9312004}}, \href {http://dx.doi.org/10.1007/BF02101598}
  {\path{doi:10.1007/BF02101598}}.

\bibitem{FR1999moduli}
V.~V. Fock and A.~A. Rosly.
\newblock {Poisson structure on moduli of flat connections on Riemann surfaces
  and r-matrix}.
\newblock {\em Am. Math. Soc. Transl.}, 191:67--86, 1999.
\newblock \href {http://arxiv.org/abs/math/9802054v2}
  {\path{arXiv:math/9802054v2}}.

\bibitem{ASS2004}
G.~Amelino-Camelia, L.~Smolin, and A.~Starodubtsev.
\newblock {Quantum symmetry, the cosmological constant and Planck-scale
  phenomenology}.
\newblock {\em Class. Quantum Gravity}, 21(13):3095--3110, 2004.
\newblock \href {http://arxiv.org/abs/hep-th/0306134}
  {\path{arXiv:hep-th/0306134}}, \href
  {http://dx.doi.org/10.1088/0264-9381/21/13/002}
  {\path{doi:10.1088/0264-9381/21/13/002}}.

\bibitem{MS2009generalized}
C.~Meusburger and B.~J. Schroers.
\newblock {Generalised Chern-Simons actions for 3d gravity and
  $\kappa$-Poincar{\'{e}} symmetry}.
\newblock {\em Nucl. Phys. B}, 806(3):462--488, 2009.
\newblock \href {http://arxiv.org/abs/0805.3318v2} {\path{arXiv:0805.3318v2}},
  \href {http://dx.doi.org/10.1016/j.nuclphysb.2008.06.023}
  {\path{doi:10.1016/j.nuclphysb.2008.06.023}}.

\bibitem{MS2009semidualization}
S.~Majid and B.~J. Schroers.
\newblock {q -deformation and semidualization in 3D quantum gravity}.
\newblock {\em J. Phys. A: Math. Theor.}, 42(42):425402, 2009.
\newblock \href {http://arxiv.org/abs/0806.2587} {\path{arXiv:0806.2587}},
  \href {http://dx.doi.org/10.1088/1751-8113/42/42/425402}
  {\path{doi:10.1088/1751-8113/42/42/425402}}.

\bibitem{BHM2014tallinn}
A.~Ballesteros, F.~J. Herranz, and F.~Musso.
\newblock {On quantum deformations of (anti-)de Sitter algebras in (2+1)
  dimensions}.
\newblock {\em J. Phys. Conf. Ser.}, 532(1):012002, 2014.
\newblock \href {http://arxiv.org/abs/1302.0684v2} {\path{arXiv:1302.0684v2}},
  \href {http://dx.doi.org/10.1088/1742-6596/532/1/012002}
  {\path{doi:10.1088/1742-6596/532/1/012002}}.

\bibitem{BGHMN2017dice}
A.~Ballesteros, I.~Guti{\'{e}}rrez-Sagredo, F.~J. Herranz, C.~Meusburger, and
  P.~Naranjo.
\newblock {Quantum groups and noncommutative spacetimes with cosmological
  constant}.
\newblock {\em J. Phys. Conf. Ser.}, 880(1):012023, 2017.
\newblock \href {http://arxiv.org/abs/1702.04704} {\path{arXiv:1702.04704}},
  \href {http://dx.doi.org/10.1088/1742-6596/880/1/012023}
  {\path{doi:10.1088/1742-6596/880/1/012023}}.

\bibitem{MercatiSergola2018constraints}
F.~Mercati and M.~Sergola.
\newblock {Physical constraints on quantum deformations of spacetime
  symmetries}.
\newblock {\em Nucl. Phys. B}, 933:320--339, 2018.
\newblock \href {http://arxiv.org/abs/1802.09483} {\path{arXiv:1802.09483}},
  \href {http://dx.doi.org/10.1016/j.nuclphysb.2018.06.014}
  {\path{doi:10.1016/j.nuclphysb.2018.06.014}}.

\bibitem{PW1990}
P.~Podle{\'{s}} and S.~L. Woronowicz.
\newblock {Quantum deformation of Lorentz group}.
\newblock {\em Commun. Math. Phys.}, 130(2):381--431, 1990.
\newblock \href {http://dx.doi.org/10.1007/BF02473358}
  {\path{doi:10.1007/BF02473358}}.

\bibitem{BM1998topological}
F.~A. Bais and N.~M. Muller.
\newblock {Topological field theory and the quantum double of SU(2)}.
\newblock {\em Nucl. Phys. B}, 530(1-2):349--400, 1998.
\newblock \href {http://arxiv.org/abs/hep-th/9804130}
  {\path{arXiv:hep-th/9804130}}, \href
  {http://dx.doi.org/10.1016/S0550-3213(98)00572-0}
  {\path{doi:10.1016/S0550-3213(98)00572-0}}.

\bibitem{BNR2002}
E.~Buffenoir, K.~Noui, and P.~Roche.
\newblock {Hamiltonian quantization of Chern-Simons theory with SL(2,R) group}.
\newblock {\em Class. Quantum Gravity}, 19(19):4953--5015, 2002.
\newblock \href {http://arxiv.org/abs/hep-th/0202121}
  {\path{arXiv:hep-th/0202121}}, \href
  {http://dx.doi.org/10.1088/0264-9381/19/19/313}
  {\path{doi:10.1088/0264-9381/19/19/313}}.

\bibitem{MN2010hilbert}
C.~Meusburger and K.~Noui.
\newblock {The Hilbert space of 3d gravity: quantum group symmetries and
  observables}.
\newblock {\em Adv. Theor. Math. Phys.}, 14(6):1651--1715, 2010.
\newblock \href {http://arxiv.org/abs/0809.2875} {\path{arXiv:0809.2875}}.

\bibitem{MS2008quaternionic}
C.~Meusburger and B.~J. Schroers.
\newblock {Quaternionic and Poisson-Lie structures in three-dimensional
  gravity: the cosmological constant as deformation parameter}.
\newblock {\em J. Math. Phys.}, 49(8):083510, 2008.
\newblock \href {http://arxiv.org/abs/0708.1507} {\path{arXiv:0708.1507}},
  \href {http://dx.doi.org/10.1063/1.2973040} {\path{doi:10.1063/1.2973040}}.

\bibitem{BHM2013cqg}
A.~Ballesteros, F.~J. Herranz, and C.~Meusburger.
\newblock {Drinfel'd doubles for (2+1)-gravity}.
\newblock {\em Class. Quantum Gravity}, 30(15):155012, 2013.
\newblock \href {http://arxiv.org/abs/1303.3080} {\path{arXiv:1303.3080}},
  \href {http://dx.doi.org/10.1088/0264-9381/30/15/155012}
  {\path{doi:10.1088/0264-9381/30/15/155012}}.

\bibitem{BHM2010plb}
A.~Ballesteros, F.~J. Herranz, and C.~Meusburger.
\newblock {Three-dimensional gravity and Drinfel'd doubles: Spacetimes and
  symmetries from quantum deformations}.
\newblock {\em Phys. Lett. B}, 687(4-5):375--381, 2010.
\newblock \href {http://arxiv.org/abs/1001.4228} {\path{arXiv:1001.4228}},
  \href {http://dx.doi.org/10.1016/j.physletb.2010.03.043}
  {\path{doi:10.1016/j.physletb.2010.03.043}}.

\bibitem{BHM2014plb}
A.~Ballesteros, F.~J. Herranz, and C.~Meusburger.
\newblock {A (2 + 1) non-commutative Drinfel'd double spacetime with
  cosmological constant}.
\newblock {\em Phys. Lett. B}, 732:201--209, 2014.
\newblock \href {http://arxiv.org/abs/1402.2884} {\path{arXiv:1402.2884}},
  \href {http://dx.doi.org/10.1016/j.physletb.2014.03.036}
  {\path{doi:10.1016/j.physletb.2014.03.036}}.

\bibitem{BHMN2014sigma}
A.~Ballesteros, F.~J. Herranz, C.~Meusburger, and P.~Naranjo.
\newblock {Twisted (2+1) $\kappa$-AdS algebra, Drinfel'd doubles and
  non-commutative spacetimes}.
\newblock {\em Symmetry, Integr. Geom. Methods Appl.}, 10:052, 26 pages, 2014.
\newblock \href {http://arxiv.org/abs/1403.4773} {\path{arXiv:1403.4773}},
  \href {http://dx.doi.org/10.3842/SIGMA.2014.052}
  {\path{doi:10.3842/SIGMA.2014.052}}.

\bibitem{BGHOS1995quasiorthogonal}
A.~Ballesteros, N.~A. Gromov, F.~J. Herranz, M.~A. del Olmo, and M.~Santander.
\newblock {Lie bialgebra contractions and quantum deformations of
  quasi-orthogonal algebras}.
\newblock {\em J. Math. Phys.}, 36(10):5916--5937, 1995.
\newblock \href {http://arxiv.org/abs/hep-th/9412083v3}
  {\path{arXiv:hep-th/9412083v3}}, \href {http://dx.doi.org/10.1063/1.531368}
  {\path{doi:10.1063/1.531368}}.

\bibitem{Dijkhuizen1994}
M.~S. Dijkhuizen and T.~H. Koornwinder.
\newblock {Quantum homogeneous spaces, duality and quantum 2-spheres}.
\newblock {\em Geom. Dedicata}, 52(3):291--315, 1994.
\newblock \href {http://dx.doi.org/10.1007/BF01278478}
  {\path{doi:10.1007/BF01278478}}.

\bibitem{BMN2017homogeneous}
A.~Ballesteros, C.~Meusburger, and P.~Naranjo.
\newblock {AdS Poisson homogeneous spaces and Drinfel'd doubles}.
\newblock {\em J. Phys. A: Math. Theor.}, 50(39):395202, 2017.
\newblock \href {http://arxiv.org/abs/1701.04902v1}
  {\path{arXiv:1701.04902v1}}, \href
  {http://dx.doi.org/10.1088/1751-8121/aa858c}
  {\path{doi:10.1088/1751-8121/aa858c}}.

\bibitem{BM2018extended}
A.~Ballesteros and F.~Mercati.
\newblock {Extended noncommutative Minkowski spacetimes and hybrid gauge
  symmetries}.
\newblock {\em Eur. Phys. J. C}, 78:615, 2018.
\newblock \href {http://arxiv.org/abs/1805.07099} {\path{arXiv:1805.07099}},
  \href {http://dx.doi.org/10.1140/epjc/s10052-018-6097-1}
  {\path{doi:10.1140/epjc/s10052-018-6097-1}}.

\bibitem{Gomez2000}
X.~Gomez.
\newblock {Classification of three-dimensional Lie bialgebras}.
\newblock {\em J. Math. Phys.}, 41(7):4939--4956, 2000.
\newblock \href {http://dx.doi.org/10.1063/1.533385}
  {\path{doi:10.1063/1.533385}}.

\bibitem{SH2002}
L.~Snobl and L.~Hlavaty.
\newblock {Classification of 6-dimensional real Drinfeld doubles}.
\newblock {\em Int. J. Mod. Phys. A}, 17(28):4043--4067, 2002.
\newblock \href {http://arxiv.org/abs/math/0202210v2}
  {\path{arXiv:math/0202210v2}}, \href
  {http://dx.doi.org/10.1142/S0217751X02010571}
  {\path{doi:10.1142/S0217751X02010571}}.

\bibitem{BMS2002}
F.~A. Bais, N.~M. Muller, and B.~J. Schroers.
\newblock {Quantum group symmetry and particle scattering in (2+1)-dimensional
  quantum gravity}.
\newblock {\em Nucl. Phys. B}, 640(1-2):3--45, 2002.
\newblock \href {http://arxiv.org/abs/hep-th/0205021}
  {\path{arXiv:hep-th/0205021}}, \href
  {http://dx.doi.org/10.1016/S0550-3213(02)00572-2}
  {\path{doi:10.1016/S0550-3213(02)00572-2}}.

\bibitem{MW1998}
H.-J. Matschull and M.~Welling.
\newblock {Quantum mechanics of a point particle in (2+1)-dimensional gravity}.
\newblock {\em Class. Quantum Gravity}, 15:2981--3030, 1998.
\newblock \href {http://arxiv.org/abs/gr-qc/9708054v2}
  {\path{arXiv:gr-qc/9708054v2}}, \href
  {http://dx.doi.org/10.1088/0264-9381/15/10/008}
  {\path{doi:10.1088/0264-9381/15/10/008}}.

\bibitem{BM2003}
E.~Batista and S.~Majid.
\newblock {Noncommutative geometry of angular momentum space U(su(2))}.
\newblock {\em J. Math. Phys.}, 44(1):107--137, 2003.
\newblock \href {http://arxiv.org/abs/hep-th/0205128}
  {\path{arXiv:hep-th/0205128}}, \href {http://dx.doi.org/10.1063/1.1517395}
  {\path{doi:10.1063/1.1517395}}.

\bibitem{Majid2005time}
S.~Majid.
\newblock {Noncommutative model with spontaneous time generation and Planckian
  bound}.
\newblock {\em J. Math. Phys.}, 46:103520, 2005.
\newblock \href {http://arxiv.org/abs/hep-th/0507271}
  {\path{arXiv:hep-th/0507271}}, \href {http://dx.doi.org/10.1063/1.2084748}
  {\path{doi:10.1063/1.2084748}}.

\bibitem{JMN2009}
E.~Joung, J.~Mourad, and K.~Noui.
\newblock {Three dimensional quantum geometry and deformed Poincar{\'{e}}
  symmetry}.
\newblock {\em J. Math. Phys.}, 50(5):052503, 2009.
\newblock \href {http://arxiv.org/abs/0806.4121} {\path{arXiv:0806.4121}},
  \href {http://dx.doi.org/10.1063/1.3131682} {\path{doi:10.1063/1.3131682}}.

\bibitem{Stachura1998}
P.~Stachura.
\newblock {Poisson-Lie structures on Poincar{\'{e}} and Euclidean groups in
  three dimensions}.
\newblock {\em J. Phys. A: Math. Gen.}, 31(19):4555--4564, 1998.
\newblock \href {http://dx.doi.org/10.1088/0305-4470/31/19/018}
  {\path{doi:10.1088/0305-4470/31/19/018}}.

\bibitem{Figueroa-OFarrill2018}
J.~M. Figueroa-O'Farrill.
\newblock {Kinematical Lie algebras via deformation theory}.
\newblock {\em J. Math. Phys.}, 59(6):061701, 2018.
\newblock \href {http://arxiv.org/abs/1711.06111v2}
  {\path{arXiv:1711.06111v2}}, \href {http://dx.doi.org/10.1063/1.5016288}
  {\path{doi:10.1063/1.5016288}}.

\bibitem{MS2003poisson}
C.~Meusburger and B.~J. Schroers.
\newblock {Poisson structure and symmetry in the Chern-Simons formulation of (2
  + 1)-dimensional gravity}.
\newblock {\em Class. Quantum Gravity}, 20(11):2193--2233, 2003.
\newblock \href {http://arxiv.org/abs/gr-qc/0301108}
  {\path{arXiv:gr-qc/0301108}}, \href
  {http://dx.doi.org/10.1088/0264-9381/20/11/318}
  {\path{doi:10.1088/0264-9381/20/11/318}}.

\bibitem{OS2018}
P.~K. Osei and B.~J. Schroers.
\newblock {Classical r-matrices for the generalised Chern-Simons formulation of
  3d gravity}.
\newblock {\em Class. Quantum Gravity}, 35(7):075006, 2018.
\newblock \href {http://arxiv.org/abs/1708.07650} {\path{arXiv:1708.07650}},
  \href {http://dx.doi.org/10.1088/1361-6382/aaaa5e}
  {\path{doi:10.1088/1361-6382/aaaa5e}}.

\bibitem{ChariPressley}
V.~Chari and A.~Pressley.
\newblock {\em {A guide to Quantum Groups}}.
\newblock Cambridge University Press, Cambridge, 1994.

\bibitem{Drinfeld1983hamiltonian}
V.~Drinfeld.
\newblock {Hamiltonian structures on Lie groups, Lie bialgebras and the
  geometric meaning of the classical Yang-Baxter equations}.
\newblock {\em Sov. Math. Dokl.}, 27:68--71, 1983.

\bibitem{Drinfeld1993}
V.~Drinfeld.
\newblock {On Poisson homogeneous spaces of Poisson-Lie groups}.
\newblock {\em Theor. Math. Phys.}, 95(2):524--525, 1993.
\newblock \href {http://dx.doi.org/10.1007/BF01017137}
  {\path{doi:10.1007/BF01017137}}.

\bibitem{Zakrzewski1997}
S.~Zakrzewski.
\newblock {Poisson Structures on the Poincar{\'{e}} Group}.
\newblock {\em Commun. Math. Phys.}, 185(2):285--311, 1997.
\newblock \href {http://arxiv.org/abs/q-alg/9602001}
  {\path{arXiv:q-alg/9602001}}, \href {http://dx.doi.org/10.1007/s002200050091}
  {\path{doi:10.1007/s002200050091}}.

\bibitem{LW2006}
J.~Lukierski and M.~Woronowicz.
\newblock {New Lie-algebraic and quadratic deformations of Minkowski space from
  twisted Poincar{\'{e}} symmetries}.
\newblock {\em Phys. Lett. B}, 633(1):116--124, 2006.
\newblock \href {http://arxiv.org/abs/hep-th/0508083v3}
  {\path{arXiv:hep-th/0508083v3}}, \href
  {http://dx.doi.org/10.1016/j.physletb.2005.11.052}
  {\path{doi:10.1016/j.physletb.2005.11.052}}.

\bibitem{BHOS1994global}
A.~Ballesteros, F.~J. Herranz, M.~A. del Olmo, and M.~Santander.
\newblock {Quantum (2+1) kinematical algebras: a global approach}.
\newblock {\em J. Phys. A: Math. Gen.}, 27(4):1283--1297, 1994.
\newblock \href {http://dx.doi.org/10.1088/0305-4470/27/4/021}
  {\path{doi:10.1088/0305-4470/27/4/021}}.
  
 \bibitem{BorowiecLukierskiTolstoy2016rmatrices}
A.~Borowiec, J.~Lukierski, V.~N.~Tolstoy.
\newblock {Quantum deformations of $D=4$ Euclidean, Lorentz, Kleinian and quaternionic $\mathfrak{o}^\star(4)$ symmetries in unified $\mathfrak{o}(4;\mathbb C)$ setting}.
\newblock {\em Phys. Lett. B}, 754:176--181, 2016.
\newblock \href {http://arxiv.org/abs/1511.03653}
  {\path{arXiv:1511.03653}}, \href
  {http://dx.doi.org/10.1016/j.physletb.2016.01.016}
  {\path{doi:10.1016/j.physletb.2016.01.016}}.

\bibitem{BorowiecLukierskiTolstoy2016rmatricesaddendum}
A.~Borowiec, J.~Lukierski, V.~N.~Tolstoy.
\newblock {Quantum deformations of $D=4$ Euclidean, Lorentz, Kleinian and quaternionic $\mathfrak{o}^\star(4)$ symmetries in unified $\mathfrak{o}(4;\mathbb C)$ setting -- Addendum}.
\newblock {\em Phys. Lett. B}, 770:426--430, 2017.
\newblock \href {http://arxiv.org/abs/arXiv:1704.06852}
  {\path{arXiv:1704.06852}}, \href
  {http://dx.doi.org/10.1016/j.physletb.2017.04.070}
  {\path{doi:10.1016/j.physletb.2017.04.070}}.

\bibitem{NappiWitten1993}
C.~R. Nappi and E.~Witten.
\newblock {Wess-Zumino-Witten model based on a nonsemisimple group}.
\newblock {\em Phys. Rev. Lett.}, 71(23):3751--3753, 1993.
\newblock \href {http://arxiv.org/abs/hep-th/9310112}
  {\path{arXiv:hep-th/9310112}}, \href
  {http://dx.doi.org/10.1103/PhysRevLett.71.3751}
  {\path{doi:10.1103/PhysRevLett.71.3751}}.

\bibitem{CJ1993extendedpoincare}
D.~Cangemi and R.~Jackiw.
\newblock {Poincar{\'{e}} gauge theory for gravitational forces in (1+1)
  dimensions}.
\newblock {\em Ann. Phys. (N. Y).}, 225:229--263, 1993.
\newblock \href {http://arxiv.org/abs/hep-th/9302026}
  {\path{arXiv:hep-th/9302026}}, \href
  {http://dx.doi.org/10.1006/aphy.1993.1058}
  {\path{doi:10.1006/aphy.1993.1058}}.

\bibitem{BCO2005quantizationDD}
A.~Ballesteros, E.~Celeghini, and M.~A. {Del Olmo}.
\newblock {Quantization of Drinfel'd doubles}.
\newblock {\em J. Phys. A: Math. Gen.}, 38:3909--3922, 2005.
\newblock \href {http://arxiv.org/abs/math/0411389}
  {\path{arXiv:math/0411389}}, \href
  {http://dx.doi.org/10.1088/0305-4470/38/18/003}
  {\path{doi:10.1088/0305-4470/38/18/003}}.

\bibitem{HS2002tdual2D}
L.~Hlavaty and L.~Snobl.
\newblock {Classification of Poisson-Lie T-dual models with two-dimensional
  targets}.
\newblock {\em Mod. Phys. Lett. A}, 17(07):429--434, 2002.
\newblock \href {http://arxiv.org/abs/hep-th/0110139}
  {\path{arXiv:hep-th/0110139}}, \href
  {http://dx.doi.org/10.1142/S0217732302006515}
  {\path{doi:10.1142/S0217732302006515}}.

\bibitem{BHOPS1995tmatrix}
A.~Ballesteros, F.~J. Herranz, M.~A. del Olmo, C.~M. Pere{\~{n}}a, and
  M.~Santander.
\newblock {Non-standard quantum (1+1) Poincare group: a T-matrix approach}.
\newblock {\em J. Phys. A: Math. Gen.}, 28:7113--7125, 1995.
\newblock \href {http://arxiv.org/abs/q-alg/9501029v1}
  {\path{arXiv:q-alg/9501029v1}}, \href
  {http://dx.doi.org/10.1088/0305-4470/28/24/012}
  {\path{doi:10.1088/0305-4470/28/24/012}}.
  
\bibitem{BHP1997nullplane}
A.~Ballesteros, F.~J. Herranz, and C.~M. Pere{\~{n}}a.
\newblock {Null-plane quantum universal R-matrix}.
\newblock {\em Phys. Lett. B}, 391(1-2):71--77, 1997.
\newblock \href {http://dx.doi.org/10.1016/S0370-2693(96)01435-9}
  {\path{doi:10.1016/S0370-2693(96)01435-9}}.

\bibitem{AF2018}
T.~Andrzejewski and J.~M. Figueroa-O'Farrill.
\newblock {Kinematical Lie algebras in 2 + 1 dimensions}.
\newblock {\em J. Math. Phys.}, 59(6):061703, 2018.
\newblock \href {http://arxiv.org/abs/1802.04048v2}
  {\path{arXiv:1802.04048v2}}, \href {http://dx.doi.org/10.1063/1.5025785}
  {\path{doi:10.1063/1.5025785}}.

\bibitem{BHN2014cqg}
A.~Ballesteros, F.~J. Herranz, and P.~Naranjo.
\newblock {From Lorentzian to Galilean (2+1) gravity: Drinfeld doubles,
  quantization and noncommutative spacetimes}.
\newblock {\em Class. Quantum Gravity}, 31:245013, 2014.
\newblock \href {http://arxiv.org/abs/1408.3689} {\path{arXiv:1408.3689}},
  \href {http://dx.doi.org/10.1088/0264-9381/31/24/245013}
  {\path{doi:10.1088/0264-9381/31/24/245013}}.

\bibitem{PS2009approach}
G.~Papageorgiou and B.~J. Schroers.
\newblock {A Chern-Simons approach to Galilean quantum gravity in 2+1
  dimensions}.
\newblock {\em J. High Energy Phys.}, 11:009, 2009.
\newblock \href {http://arxiv.org/abs/0907.2880} {\path{arXiv:0907.2880}},
  \href {http://dx.doi.org/10.1088/1126-6708/2009/11/009}
  {\path{doi:10.1088/1126-6708/2009/11/009}}.

\bibitem{PS2010}
G.~Papageorgiou and B.~J. Schroers.
\newblock {Galilean quantum gravity with cosmological constant and the extended
  q-Heisenberg algebra}.
\newblock {\em J. High Energy Phys.}, 11:020, 2010.
\newblock \href {http://arxiv.org/abs/1008.0279v1} {\path{arXiv:1008.0279v1}},
  \href {http://dx.doi.org/10.1007/JHEP11(2010)020}
  {\path{doi:10.1007/JHEP11(2010)020}}.

\bibitem{BHN2015towards31}
A.~Ballesteros, F.~J. Herranz, and P.~Naranjo.
\newblock {Towards (3+1) gravity through Drinfel'd doubles with cosmological
  constant}.
\newblock {\em Phys. Lett. B}, 746:37--43, 2015.
\newblock \href {http://arxiv.org/abs/1502.07518} {\path{arXiv:1502.07518}},
  \href {http://dx.doi.org/10.1016/j.physletb.2015.04.041}
  {\path{doi:10.1016/j.physletb.2015.04.041}}.

\bibitem{BHMN2017kappa3+1}
A.~Ballesteros, F.~J. Herranz, F.~Musso, and P.~Naranjo.
\newblock {The $\kappa$-(A)dS quantum algebra in (3 + 1) dimensions}.
\newblock {\em Phys. Lett. B}, 766:205--211, 2017.
\newblock \href {http://arxiv.org/abs/1612.03169} {\path{arXiv:1612.03169}},
  \href {http://dx.doi.org/10.1016/j.physletb.2017.01.020}
  {\path{doi:10.1016/j.physletb.2017.01.020}}.

\bibitem{Figueroa-OFarrill2018higher}
J.~M. Figueroa-O'Farrill.
\newblock {Higher-dimensional kinematical Lie algebras via deformation theory}.
\newblock {\em J. Math. Phys.}, 59(6):061702, 2018.
\newblock \href {http://arxiv.org/abs/1711.07363v2}
  {\path{arXiv:1711.07363v2}}, \href {http://dx.doi.org/10.1063/1.5016616}
  {\path{doi:10.1063/1.5016616}}.

\bibitem{KS1995duality}
C.~Klim{\v{c}}{\'{i}}k and P.~{\v{S}}evera.
\newblock {Dual non-Abelian duality and the Drinfeld double}.
\newblock {\em Phys. Lett. B}, 351(4):455--462, 1995.
\newblock \href {http://arxiv.org/abs/hep-th/9502122}
  {\path{arXiv:hep-th/9502122}}, \href
  {http://dx.doi.org/10.1016/0370-2693(95)00451-P}
  {\path{doi:10.1016/0370-2693(95)00451-P}}.

\bibitem{Sfetsos1998}
K.~Sfetsos.
\newblock {Poisson-Lie T-duality beyond the classical level and the
  renormalization group}.
\newblock {\em Phys. Lett. B}, 432(3-4):365--375, 1998.
\newblock \href {http://arxiv.org/abs/hep-th/9803019}
  {\path{arXiv:hep-th/9803019}}, \href
  {http://dx.doi.org/10.1016/S0370-2693(98)00666-2}
  {\path{doi:10.1016/S0370-2693(98)00666-2}}.

\bibitem{LledoVaradarajan1998}
M.~A. Lled{\'{o}} and V.~S. Varadarajan.
\newblock {SU(2) Poisson-Lie T duality}.
\newblock {\em Lett. Math. Phys.}, 45:247--257, 1998.
\newblock \href {http://dx.doi.org/10.1023/A:1007498803198}
  {\path{doi:10.1023/A:1007498803198}}.

\bibitem{JafarizadehRezaei1999}
M.~A. Jafarizadeh and A.~Rezaei-Aghdam.
\newblock {Poisson-Lie T-duality and Bianchi type algebras}.
\newblock {\em Phys. Lett. B}, 458(4):477--490, 1999.
\newblock \href {http://arxiv.org/abs/hep-th/9903152}
  {\path{arXiv:hep-th/9903152}}, \href
  {http://dx.doi.org/10.1016/S0370-2693(99)00571-7}
  {\path{doi:10.1016/S0370-2693(99)00571-7}}.

\bibitem{BeggsMajid2001tduality}
E.~J. Beggs and S.~Majid.
\newblock {Poisson-Lie T-duality for quasitriangular Lie bialgebras}.
\newblock {\em Commun. Math. Phys.}, 220(3):455--488, 2001.
\newblock \href {http://arxiv.org/abs/math/9906040}
  {\path{arXiv:math/9906040}}, \href {http://dx.doi.org/10.1007/s002200100463}
  {\path{doi:10.1007/s002200100463}}.

\bibitem{HS2004tplurality3D}
L.~Hlavaty and L.~Snobl.
\newblock {Poisson-Lie T-plurality of three-dimensional conformally invariant
  sigma models}.
\newblock {\em J. High Energy Phys.}, JHEP05(2004)010.
\newblock \href {http://arxiv.org/abs/hep-th/0408126}
  {\path{arXiv:hep-th/0408126}}, \href
  {http://dx.doi.org/10.1088/1126-6708/2004/05/010}
  {\path{doi:10.1088/1126-6708/2004/05/010}}.
  



\end{thebibliography}
\end{document}